\documentclass[aps,amsmath,twocolumn,amssymb,floatfixng,showpacs,
superscriptaddress,footinbib]{revtex4-2}
\usepackage{bm}
\usepackage{graphicx}
\usepackage{graphics}
\usepackage{mathtools}
\usepackage{amsmath}
\usepackage{newtxtext} 
\usepackage{epstopdf}
\usepackage{hyperref}
\usepackage{hyperref}
\hypersetup{
    colorlinks,%
    citecolor=blue,%
    linkcolor=blue,%
    urlcolor=blue
}
\usepackage{tikz}\usetikzlibrary{petri}
\usepackage{soul}

\begin{document}

\newcommand{\be}   {\begin{equation}}
\newcommand{\ee}   {\end{equation}}
\newcommand{\ba}   {\begin{eqnarray}}
\newcommand{\ea}   {\end{eqnarray}}
\newcommand{\ve}  {\varepsilon}

\title{ Density waves in low-pressure bilayer nickelates }

\author{Lauro B. Braz}
\affiliation{
Instituto de F\'{\i}sica, Universidade de S\~ao Paulo, Rua do Mat\~ao 1371, S\~ao Paulo, S\~ao Paulo 05508-090, Brazil
}
\affiliation{
Theoretische Physik III, Ruhr-Universit\"at Bochum, D-44780, Bochum, Germany
}
\author{Steffen B\"otzel}
\affiliation{
Theoretische Physik III, Ruhr-Universit\"at Bochum, D-44780, Bochum, Germany
}
\author{Frank Lechermann}
\affiliation{
Theoretische Physik III, Ruhr-Universit\"at Bochum, D-44780, Bochum, Germany
}
\author{Igor Plokhikh}
\affiliation{PSI Center for Neutron and Muon Sciences CNM, 5232 Villigen PSI, Switzerland}
\affiliation{TU Dortmund University, Department of Physics, Dortmund, 44227, Germany}
\author{Rustem Khasanov}
\affiliation{PSI Center for Neutron and Muon Sciences CNM, 5232 Villigen PSI, Switzerland}
\author{Luis G.~G.~V. Dias da Silva}
\affiliation{
Instituto de F\'{\i}sica, Universidade de S\~ao Paulo, Rua do Mat\~ao 1371, S\~ao Paulo, S\~ao Paulo 05508-090, Brazil
}
\author{Ilya M. Eremin}
\affiliation{
Theoretische Physik III, Ruhr-Universit\"at Bochum, D-44780, Bochum, Germany
}

\begin{abstract}
The low-pressure phase diagram of La$_3$Ni$_2$O$_7$ provides an important reference for understanding its pressure-induced high-temperature superconductivity. 
While the spin-density-wave transition at $T_{\text{SDW}}\approx150$ K is increasingly well established, the origin of the second density-wave transition at $T_{\text{DW}}\approx130$ K has remained unresolved. 
Here, we perform unrestricted Hartree–Fock calculations to investigate the potential origin of the second transition. {Within the orthorhombic phase, the degeneracy between possible ordering wavevectors at $\boldsymbol{Q}_{Y}=(0,\pi)$ and at $\boldsymbol{Q}_{X}=(\pi,0)$ is lifted and the electronic system} develops a double-stripe spin-density wave with ordering vector $\boldsymbol{Q}_{Y}=(0,\pi)$.
{We identify that the pure double stripe spin state is unstable in La$_3$Ni$_2$O$_7$ towards a commensurate charge-density wave instability, which favors a spin-modulated double stripe order with intertwined charge and spin instabilities and establish the hierarchy of ordered states in La$_3$Ni$_2$O$_7$. We further discuss our results in the context of available experimental literature and propose further experimental tests to elucidate the origin of the SDW/DW states in this system.}
\end{abstract}

\maketitle

\section{Introduction}
\label{sec:Intro}
Nickelate superconductors represent one of the most significant recent advances in the study of high-$T_c$ superconductivity. 
Under elevated pressure, La$_3$Ni$_2$O$_7$ exhibits a structural phase transition from the \textit{Amam} (orthorhombic) to the \textit{I4/mmm} (tetragonal) space group and becomes superconducting with a critical temperature of up to 96 K under partial rare-earth substitution \cite{sunSignaturesSuperconductivity802023,Li_SuperconductivityBilayerNickelateLa3Ni2O7_2025,Li_BulkSuperconductivityUpTo96KInPressurizedNickelateSingleCrystals_2025}.
At the outset, it was believed that the distinct low- and high-pressure crystal structures are directly responsible for the superconducting state \cite{sunSignaturesSuperconductivity802023}.
However, experimental investigations have reported that samples that have a tetragonal structure at ambient pressure {do not show the spin density wave transition and are not superconducting under pressure. This indicates that instead of the crystal structure alone, the density waves found at lower pressures are also important to superconductivity, that appears at higher pressures} \cite{Shi2025SpinDensityWaveNickelate,Fan2026EvolutionNickelate}.
Similar findings have been reported for the trilayer nickelate \cite{Shi2025AbsenceSuperconductivityLa4Ni3O10}.
Therefore, a deeper understanding of the low-pressure phase diagram of La$_3$Ni$_2$O$_7$ is essential to elucidate the nature of the high-pressure superconducting state.

At ambient pressure, the phase diagram of the bilayer nickelate displays a spin-density wave (SDW) at a critical temperature $T_{\text{SDW}}\sim150$ K followed by another density wave-like order that sets in at $T_{\text{DW}}\sim130$ K \cite{Puphal2026NickelateReview,Khasanov_PressureEnhancedSplittingDensityWaveTransitions_2025,Zhou2025PressureSDWNickelate,Zhang2024HTSCNickelate,Wu2025UltrafastDensityWaves,Wang2024PressureSuperconductivityNickelate,Khasanov2026IsotopeNickelate,Liu2026DensityWavePhases}, which origin is still unclear.
With increasing pressure, the density wave (DW) is suppressed, while the SDW is slightly enhanced~\cite{Puphal2026NickelateReview,Khasanov_PressureEnhancedSplittingDensityWaveTransitions_2025,Zhou2025PressureSDWNickelate,Khasanov2026IsotopeNickelate,Liu2026DensityWavePhases}.
The dissonant behavior between SDW and DW in the bilayer system is in sharp contrast to the incommensurate SDW and charge-density wave (CDW) in the trilayer nickelate La$_4$Ni$_3$O$_{10}$, in which the SDW and CDW are intertwined and are equally suppressed by pressure \cite{Puphal2026NickelateReview,Xu2025CollapseDensityWaveNickelate,Li2025UnidirectionalCDWLa4Ni3O10,Khasanov2026PressureIsotopeLa4Ni3O10,Norman2025LandauNickelate}.
Although recent experiments have accumulated evidence that the DW in La$_3$Ni$_2$O$_7$ is likely a CDW \cite{Puphal2026NickelateReview,Liu2023ChargeSpinNickelate,Kakoi_MultibandMetallicGroundStateNickelates_2024,Luo_MicroscopicEvidenceChargeSpinDensityWaves_2025,Khasanov2026IsotopeNickelate}, the CDW in the trilayer has somewhat different origin from the DW in the bilayer compound.

At low pressures, La$_3$Ni$_2$O$_7$ displays a multiband metallic character at all temperatures, including within the ordered phases \cite{Puphal2026NickelateReview,Zhang1994LayeredNickelate,Liu2023ChargeSpinNickelate,Ling2000NeutronDiffractionNickelate,Chen2024ElectronicMagneticNickelate,Kakoi_MultibandMetallicGroundStateNickelates_2024,Fan2024TunnelingNickelate,Chen2025MultibandNickelate,Michon2026MetallicCrossoverNickelate}.
This makes the low-pressure bilayer nickelate an itinerant magnet.
Moreover, there is no experimental evidence of a structural phase transition below {$T_{\text{SDW}}$} \cite{Ling2000NeutronDiffractionNickelate,Liu2023ChargeSpinNickelate,Xie_StrongInterlayerMagneticExchangeCoupling_2024,Gupta2025SpinStripeNickelate,He2026AnisotropicNickelate,Misawa2026,Reiss2026,Liu2026DensityWavePhases,Sasaki2026StructuralPhaseDiagram}.
Instead, an electronic unidirectional and commensurate CDW was reported \cite{Luo_MicroscopicEvidenceChargeSpinDensityWaves_2025}. 
In their turn, supported by spin models, experimental magnetic probes provide evidence of a double-stripe order for the SDW with a magnetic wavevector $\boldsymbol{Q}_Y=(0,\pi)$, {where the antiferromagnetic order takes place along the longer lattice parameter $b>a$ }\cite{Plokhikh2025SpinDensityNickelate,Gupta2025SpinStripeNickelate,Zhou2025PressureSDWNickelate,Yashima2025SpinStripeNickelate,Xie_StrongInterlayerMagneticExchangeCoupling_2024}.
From the theory side, normal-state DFT+DMFT calculations show good agreement with ARPES data \cite{Leonov2025SpinChargeStripesNickelate,Liu2025OrbitalSelectivityNickelate,Wang2024ElectronicMagneticNickelate}.
However, both DFT and the strongly correlated corrections in DFT+DMFT predict that inside the ordered phase, the system is insulating \cite{Yi2024ChargeDensityWaveNickelate,LaBollita2024SDWNickelate,Ni2025SpinDensityWaveNickelate,Leonov2025SpinChargeStripesNickelate,Chen2025ChargeSpinInstabilities}.
Accounting for the SDW, the proposed phases yielding such insulating regimes include structural CDW proposals \cite{Yi2024ChargeDensityWaveNickelate,Chen2025ChargeSpinInstabilities,Yi2025StrainPressureNickelate}, pure SDW \cite{Zhang2024StructuralNickelate,LaBollita2024SDWNickelate,Wang2025SpinStripesNickelate,Liu2025DiagonalStripeNickelate}, and an electronic CDW with a large charge imbalance of Ni$^{3+}$ and Ni$^{2+}$ in the NiO$_6$ plane \cite{Leonov2025SpinChargeStripesNickelate,Ni2025SpinDensityWaveNickelate,Chen2025ChargeSpinInstabilities,Yi2025StrainPressureNickelate}, most of which predict the SDW as a double-stripe order.
In the above cases where the CDW and SDW are predicted together, the critical temperature of both orders coincide \cite{Wang2025SpinStripesNickelate,Mei2026,Zhan2026,Deng2026IntersiteCoulomb}.
In addition, the $\mu$SR \cite{Chen2024SpinDensityWaves,Khasanov_PressureEnhancedSplittingDensityWaveTransitions_2025} and nuclear quadrupole resonance (NQR) \cite{Yashima2025SpinStripeNickelate} measurements were interpreted in terms of a spin-spinless double stripe order.

Resonant soft x-ray scattering, polarimetry, and polarized ultrafast spectroscopy measurements shed new light to the understanding of the low-pressure phase diagram of La$_3$Ni$_2$O$_7$ suggesting the spin-density wave features stripes that break $C_4$ rotational symmetry, resembling the nematic state of Fe pnictides \cite{Gupta2025SpinStripeNickelate,Wu2026ElectronicNematicityNickelate,Wu2025UltrafastDensityWaves}.
This is an expected behavior for the present orthorhombic structure, which naturally breaks $C_4$ symmetry.
In this work, motivated by the abundance of experimental probes, especially neutron scattering experiments \cite{Xie_StrongInterlayerMagneticExchangeCoupling_2024,Zhou2026SpinFluctuationsNickelates,Chen2026NatureMagnetismNickelate}, we explore the $\boldsymbol{Q}_Y=(0,\pi)$ magnetic order in low-pressure La$_3$Ni$_2$O$_7$ {within a multi-orbital Hubbard-Hund Hamiltonian}.
We consider both magnetic and charge order parameters and solve the unrestricted Hartree-Fock problem to probe for orders in both channels.
Our findings indicate a metallic double-stripe SDW emerging at $T_{\text{SDW}}$.
At a temperature $T_{\text{DW}}<T_{\text{SDW}}$, a unidirectional CDW emerges breaking in-plane site equivalency in the unit cell together with a redistribution of magnetic moments, generating low-spin sites.
The charge order parameter is one order of magnitude smaller than the magnetic order parameter, which may explain the experimental difficulty in its observation using thermodynamic probes. 
Note, our calculations result in the ratio $T_{\text{DW}}/T_{\text{SDW}}\sim0.6$ at ambient pressure.
{At the same time, the charge order transition temperature, and thus the ratio of $T_{\text{DW}}/T_{\text{SDW}}$, depend sensitively on the magnitude of the crystal field splitting, which could explain the pressure-dependent evolution of $T_{\text{DW}}$, found experimentally.}
We analyze our finding in the context of the recent experiments including the $\mu$SR data.

\begin{figure}[t]
\begin{center}
\includegraphics[width=1.0\columnwidth]{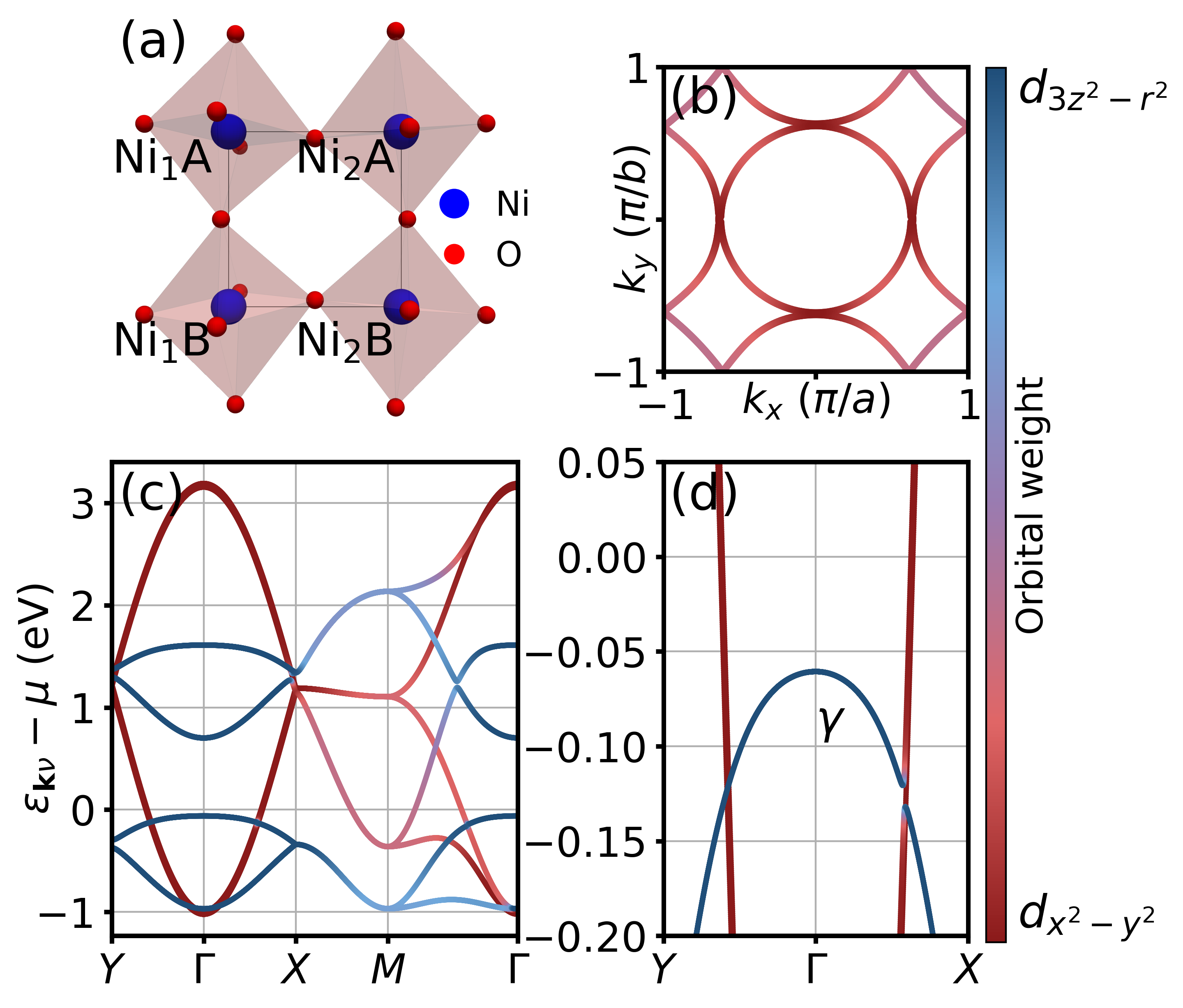}
\caption{ The unit cell in the orthorhombic crystal structure relaxed from DFT calculations is shown and the site/layer notation is defined in (a).
Straight lines connect the Ni atom centers, evidencing the tilted oxygen geometry.
The normal-state Fermi surface (b) and electronic band structures (c,d), as obtained self-consistently for $T/T_{\text{SDW}}>1$, are shown.
Colors denote the $d_{3z^2-r^2}$ (blue) and $d_{x^2-y^2}$ (red) spectral weights and opacity quantifies the total spectral weight at each band and momentum tuple.
In (b), momentum is given in terms of the lattice constants $a\approx0.539$ nm and $a/b\approx0.99$.
In panel (d), we zoom-in the $\gamma$ pocket, which is pushed below the Fermi level in the Hartree-Fock cycles \cite{Wang2024ElectronicMagneticNickelate,Yang_OrbitalDependentElectronCorrelationLa3Ni2O7_2024}.
}
\label{fig:model_system}
\end{center}
\end{figure}

\section{Results}
\label{sec:results}
\subsection{Theory of coupled magnetic and charge orders}
\label{sec:modelmethods}
We model the non-interacting electrons in the orthorhombic crystal structure by a 8-orbital tight-binding model fitted from a DFT calculation given in Ref.~\cite{Wang2024ElectronicMagneticNickelate}.
In this model, the unit cell has all four nonequivalent Ni sites in the orthorhombic unit cell that is, two layers A and B, and two in-plane sites Ni$_1$ and Ni$_2$.
In each one of these sites, two orbitals $d_{3z^2-r^2}$ and $d_{x^2-y^2}$ are modeled, which gives rise to  eight bands in total.
The tight-binding Hamiltonian reads
\begin{align}
H_0
=
\sum_{\boldsymbol{k}}
\sum_{\mu\eta}
\sum_{\sigma}
\xi^{\mu\eta}_{\boldsymbol{k}}
\, c^{\dagger}_{\boldsymbol{k}\mu\sigma}
c_{\boldsymbol{k}\eta\sigma}, \label{eq:H0_sublattice}
\end{align}
where $\xi^{\mu\eta}_{\boldsymbol{k}}=\epsilon^{\mu\eta}_{\boldsymbol{k}}-\mu\delta_{\mu\eta}$, in which we define the chemical potential $\mu$ and the tight-binding matrix $\epsilon^{\mu\eta}_{\boldsymbol{k}}$ that includes hoppings up to third-nearest neighbors and a tunning parameter $\Delta_{xz}=\epsilon_{d_{x^2}}-\epsilon_{d_{z^2}}$ defined as the difference between onsite energies {(crystal-field splitting)} for $d_{x^2-y^2}$ and $d_{3z^2-r^2}$ orbitals.
Here, $c_{\boldsymbol{k}\mu\sigma}$ annihilates an electron with momentum $\boldsymbol{k}$ and spin $\sigma$ in orbital $\mu$ indices.
The pressure evolution is estimated from the crystal-field splitting data as calculated by DFT from the experimental crystal structures of Ref.~\cite{Xu2026} and is given in Table \ref{tab:cf} of Appendix~\ref{appendix:computational}.

Fig. \ref{fig:model_system}(a) shows the orthorhombic unit cell, in which the oxygen octahedra around Ni sites are tilted, as observed in $Amam$-symmetric samples \cite{Bhatt2025StructuralOriginsNickelate}.
The orbital-dependent spectral weights are shown in panels (b-d).
The normal-state Fermi surface is shown in panel (b), where we note that the choice of a two-atom unit cell distorts the Brillouin zone.
Panel (c) shows the band structure along high-symmetry directions and a zoomed-in view is shown in panel (d),
in which a pocket dubbed $\gamma$ centered at the $\boldsymbol{\Gamma}$ high-symmetry point is highlighted.
As previously reported in hybrid functional DFT \cite{Wang2024ElectronicMagneticNickelate} and DFT+$U$ calculations \cite{Yang_OrbitalDependentElectronCorrelationLa3Ni2O7_2024}, the Hartree-Fock self-consistent cycle with finite Hubbard $U$ compensates this effect resulting in a $\gamma$ pocket {shifted} $\sim60$ meV below the Fermi level in the present calculation.
We remark that the band structure features a double degeneracy of bands at the $\boldsymbol{X}=(\pi,0)$ high-symmetry point, which is not present in the $\boldsymbol{Y}=(0,\pi)$ point, where splittings make all eight bands non-degenerate.

We consider the interactions present in a Slater-Kanamori Hamiltonian, in which onsite direct and exchange terms are considered, namely the onsite Hubbard $U$ and its interorbital component $U'$, and Hund's $J_H$ and a pair-hopping component $J_H'$.
Spin-rotational invariance imposes constraints $U'=U-2J_H$ and $J_H'=J_H$ \cite{Dagotto2001ColossalMagnetoresistance,graserNeardegeneracySeveralPairing2009}.
The Slater-Kanamori Hamiltonian can be conveniently rewritten in spin space in terms of charge and spin components, and in momentum space reads \cite{graserNeardegeneracySeveralPairing2009,Braz2024ChargeSpinFluctuations,Braz2025InterlayerInteractions}
\begin{equation}
\begin{split}
H_{\mathrm{int}}
=
\frac{1}{2}
\sum_{\boldsymbol{q}}
\sum_{\mu\eta\alpha\beta}
\Big[
&(U_c)^{\mu\eta}_{\alpha\beta}
n_{-\boldsymbol{q}\mu\eta}
n_{\boldsymbol{q}\alpha\beta} \\
-
&(U_s)^{\mu\eta}_{\alpha\beta}
\boldsymbol{S}_{-\boldsymbol{q}\mu\eta}
\cdot
\boldsymbol{S}_{\boldsymbol{q}\alpha\beta}
\Big],
\label{eq:Kanamori_nc_form}
\end{split}
\end{equation}
where the electron density $n_{\boldsymbol{q}\mu\eta}
=
\sum_{\boldsymbol{k}\sigma}
c^{\dagger}_{(\boldsymbol{k}+\boldsymbol{q})\mu\sigma}
c_{\boldsymbol{k}\eta\sigma}$
and spin density
$\boldsymbol{S}_{\boldsymbol{q}\mu\eta}
=
\frac{1}{2}
\sum_{\boldsymbol{k}\sigma\sigma'}
c^{\dagger}_{(\boldsymbol{k}+\boldsymbol{q})\mu\sigma}
\boldsymbol{\sigma}_{\sigma\sigma'}
c_{\boldsymbol{k}\eta\sigma'}$
are defined.
$\boldsymbol{\sigma}$ denotes the Pauli vector.
For brevity, we give the matrices $\hat U_s$ and $\hat U_c$ in Appendix \ref{appendix:action}.

In Appendix \ref{appendix:action}, we derive the free energy for a multiorbital system with charge $\phi_\mu$ and spin $\boldsymbol M_\mu$ order parameters from the action formalism.
The mean-field self-consistent parameters are related to the charge and spin densities by the action saddle-point solution of the multiorbital system,
\begin{align}
\phi_{\mu}
&=
-\sum_{\eta}
(U_c)^{\mu\mu}_{\eta\eta}
\langle n_{\eta} \rangle,
\\
\boldsymbol{M}_{\mu}
&=
\sum_{\eta}
(U_s)^{\mu\mu}_{\eta\eta}
\langle \boldsymbol{S}_{\eta} \rangle,
\end{align}
where $n_{\eta}\equiv n_{\boldsymbol{0}\eta\eta}$ is defined at $\boldsymbol{q}=\boldsymbol 0$ and $\boldsymbol{S}_{\eta}\equiv\boldsymbol{S}_{\boldsymbol Q\eta\eta}$ at $\boldsymbol{q}=\boldsymbol{Q}$ the magnetic ordering vector.
We only consider saddle-point solutions which are diagonal in orbital space.
Moreover, $\boldsymbol{M}_{\mu}$ is chosen to be collinear to the $\sigma_z$ direction and denoted $M_\mu\equiv M_\mu^z$.
We remark that $\phi_\mu$ may not break lattice symmetries, still its multiorbital elements can break symmetries inside the unit cell.
In this way, differences $\phi_\mu-\phi_\eta$ can constitute a charge order parameter if they involve different lattice sites or layers [see Fig.~\ref{fig:model_system}(a)].

Motivated by the orthorhombic structure of La$_3$Ni$_2$O$_7$, which naturally breaks $C_4$ symmetry, and for the compelling experimental evidence for the $(0,\pi)$ order in the system \cite{Xie_StrongInterlayerMagneticExchangeCoupling_2024,Chen2024ElectronicMagneticNickelate,Zhou2026SpinFluctuationsNickelates,Chen2026NatureMagnetismNickelate}, we choose $\boldsymbol{Q}_{Y}=(0,\pi)$. {Note, assuming ${\boldsymbol{Q}_X}=(\pi,0)$ yields much smaller $T_\text{SDW}$ and smaller gain in the free energy. 
This is also evident from our normal state calculations of the spin susceptibility, which shows $\chi_S (\boldsymbol{Q}_Y,0)>\chi_S (\boldsymbol{Q}_X,0)$ and the natural tendency of the electronic subsystem towards $\boldsymbol{Q}_{Y}=(0,\pi)$ SDW state \cite{Botzel2026}. }
The mean-field Hamiltonian matrix then reads
\begin{equation}
\hat h^{\sigma}(\boldsymbol k)=
\begin{pmatrix}
\hat\xi_{\boldsymbol k}+\hat\phi & -\sigma \hat{M} \\
-\sigma\hat M & \hat\xi_{\boldsymbol k+\boldsymbol Q}+\hat\phi
\end{pmatrix},\label{eq:HMF}
\end{equation}
which is diagonal in spin space.
This Hamiltonian matrix is diagonalized as $\hat h^{\sigma}(\boldsymbol k)|\boldsymbol{k}\nu\sigma\rangle=\epsilon_{\boldsymbol k\nu}|\boldsymbol{k}\nu\sigma\rangle$.
The antiferromagnetic order results in spin up and down degenerate solutions, characteristic of antiferromagnetism.
We also note that a charge order realized through the self-consistent parameter $\hat\phi$ is possible at the ordering vector $\boldsymbol{Q}_c=2
\boldsymbol{Q}_Y=0$.

\begin{figure}[t]
\begin{center}
\includegraphics[width=1.0\columnwidth]{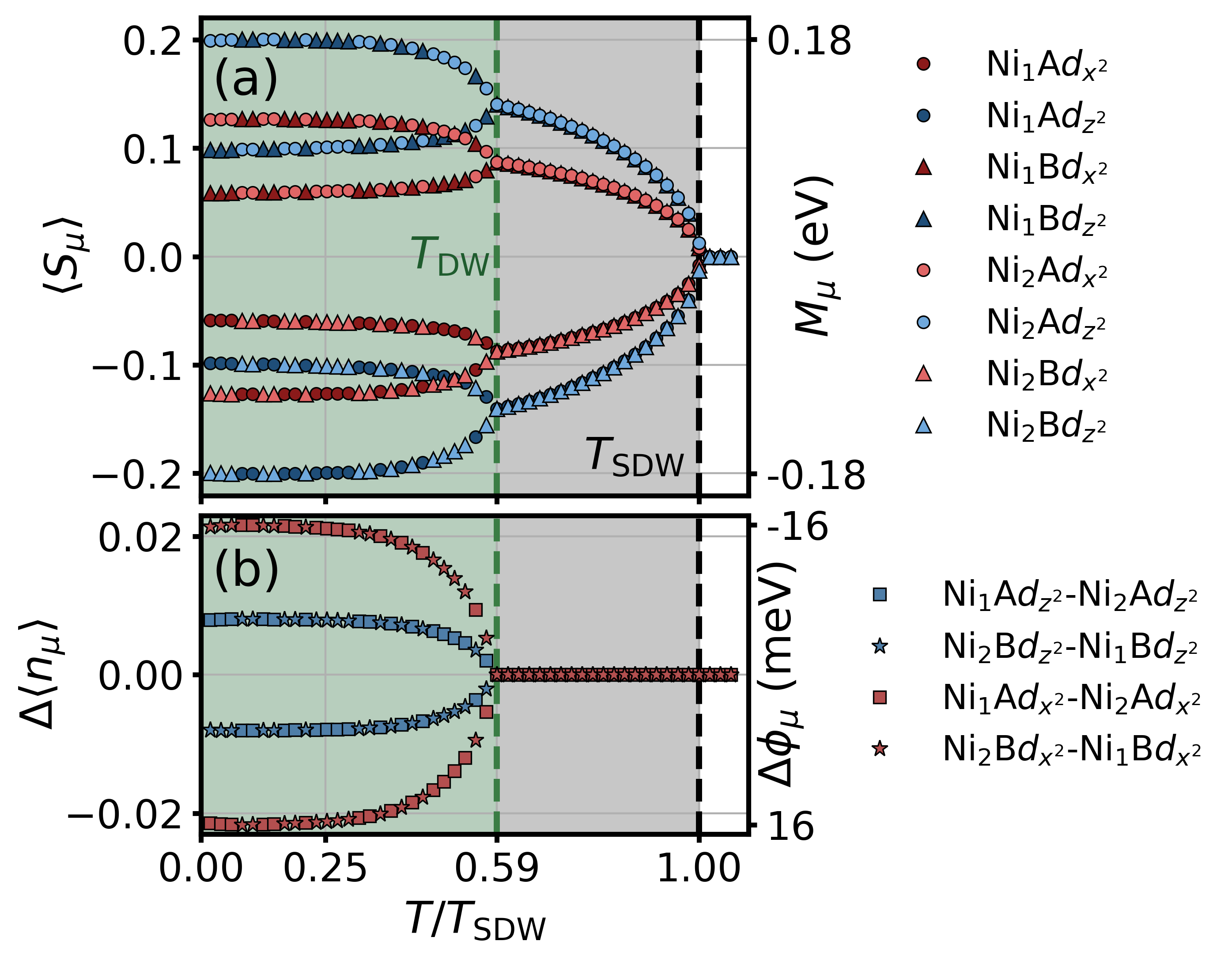}
\caption{ Spin density (a) and difference in charge densities (b) as a function of temperature at ambient pressure for the $\boldsymbol{Q}_{Y}=(0,\pi)$ magnetic ordering vector.
In panel (a), following the colors of Fig. \ref{fig:model_system}(b,c,d), blue denotes the $d_{3z^2-r^2}$ orbital and red the $d_{x^2-y^2}$.
Darker and lighter colors denote the in-plane Ni sites and symbols denote layers A (circle) and B (triangle), following the notation convention shown in Fig. \ref{fig:model_system}(a).
Panel (b) mixes sites and only keeps the color convention for orbitals.
Vertical dashed lines mark the spin-density wave (black) and density wave (green) critical temperatures.
}
\label{fig:order_parameters}
\end{center}
\end{figure}

Finally, the mean-field approximation results in the free-energy functional
\begin{equation}
\begin{split}
F
&=
\frac{1}{2}
\sum_{\mu\eta}
\phi_{\mu}
(U_c^{-1})^{\mu\mu}_{\eta\eta}
\phi_{\eta}
+
\frac{1}{2}
\sum_{\mu\eta}
M_{\mu}
(U_s^{-1})^{\mu\mu}_{\eta\eta}
M_{\eta}\\
&-
\frac{1}{\beta}
\sum_{\boldsymbol k\in RBZ}
\sum_{\eta\sigma}
\ln\left(
1+e^{-\beta (\epsilon_{\boldsymbol k\eta}-\mu)}
\right).\label{eq:FMF_Kanamori}
\end{split}
\end{equation}
Momentum summation must be carried over a reduced Brillouin zone (RBZ) to avoid double counting when doubling the unit cell size to include band folding of the magnetic ordering vector $\boldsymbol{Q}_{Y}$.
See Appendix \ref{appendix:action} for more details.
Moreover, $\beta=1/T$ is the inverse temperature.
We solve a numerical minimization problem for the grand potential $\Omega=F+\mu n$ for fixed density $n=1.5$ per Ni layer \cite{Zhang2023ElectronicStructureNickelate} in terms of the charge $\phi_\mu$ and spin $M_\mu$ order parameters, and the chemical potential $\mu$.
Also, we probe for the different solutions of the mean-field equations as a function of $U$ and $J_H$ in Appendix \ref{appendix:details} and set $U=0.61$ eV, $J_H=U/10$ throughout the main text.
This choice of parameters is motivated by $(i)$ the experimental evidence for a SDW gap of $2\Delta\sim65-100$ meV \cite{Liu2024ElectronicCorrelationsNickelate,Meng2024DensityWaveNickelate,Fan2024TunnelingNickelate,Wu2025UltrafastDensityWaves,Zhu2026HighEnergyExcitationsNickelate,Kakoi_MultibandMetallicGroundStateNickelates_2024} and $(ii)$ the metallic character of the bilayer nickelate even at low temperatures and especially probed by ARPES measurements \cite{Zhang1994LayeredNickelate,Liu2023ChargeSpinNickelate,Ling2000NeutronDiffractionNickelate,Chen2024ElectronicMagneticNickelate,Kakoi_MultibandMetallicGroundStateNickelates_2024,Fan2024TunnelingNickelate,Chen2025MultibandNickelate,Yang_OrbitalDependentElectronCorrelationLa3Ni2O7_2024,Li2024PseudogapNickelate}.

\begin{figure}[t]
\begin{center}
\includegraphics[width=1.0\columnwidth]{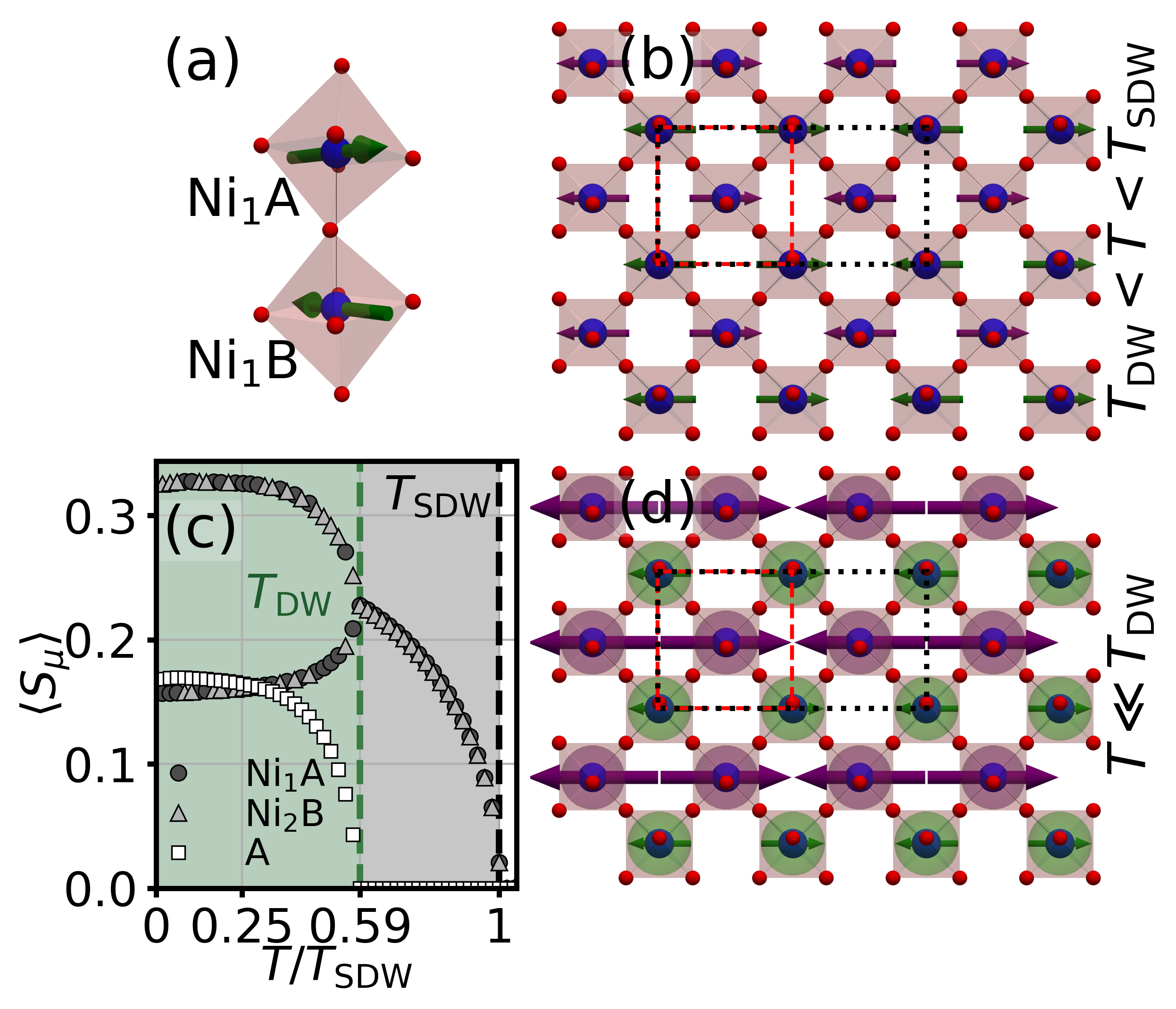}
\caption{ Site-resolved (trace over orbitals) spin and charge-modulation expectation values.
Panel (a) shows the antiferromagnetic alignment between layers.
Top view of the bilayer structure is depicted in panels (b,d) for the SDW and SDW+CDW phases, respectively.
Normal-state (red dashed lines) and magnetic (black dotted lines) unit cells are shown.
Ni and O atom colors follow convention of Fig.~\ref{fig:model_system}.
Green and purple colors for arrows (spin density) and spheres (deviations from average charge density) indicate distinct sublattice sites Ni$_1$ and Ni$_2$, respectively.
The charge modulation features opposite sign in the in-plane Ni$_1$ and Ni$_2$ sites.
Panel (c) shows the total spin expectation value (summed over orbitals) in the sublattice sites Ni$_1$A (dark gray circles), Ni$_2$B (light gray triangles), and layer A (Ni$_1$A plus Ni$_2$A, white squares).
}
\label{fig:orders}
\end{center}
\end{figure}


\subsection{Density waves}
\label{sec:dw}
There are two known characteristic temperatures related to the density waves in the low-pressure bilayer nickelates, the spin-density wave critical temperature $T_{\text{SDW}}$ and the density wave onset $T_{\text{DW}}$, where a hierarchy between these temperatures such that $T_{\text{SDW}}>T_{\text{DW}}$ is well known.
Here, we probe for both orders and show that this hierarchy features a dominant SDW order parameter ($\sim180$ meV) with a low-energy CDW energy scale ($\sim16$ meV) at ambient pressure.

Associated with the temperature $T_{\text{SDW}}$ (above $T_{\text{DW}}$), we report a double-stripe spin-density wave, as seen in the light gray region of Fig.~\ref{fig:order_parameters}(a), which shows the average spin density $\langle S_\mu\rangle$ as a function of temperature. 
This is a realization of the $\boldsymbol{Q}_{Y}=(0,\pi)$ order.
The site (orbital averaged) spin densities are shown in Fig.~\ref{fig:orders}(a) (side view) and (b) (top view), where arrow colors denote the in-plane sublattice sites Ni$_1$ (green) and Ni$_2$ (purple).
The double-stripe pattern is visualized in panel (b).
The antiferromagnetic state is orbital dependent, showing a splitting of gap amplitudes in Fig.~\ref{fig:order_parameters}(a) for the $d_{3z^2-r^2}$ (blue color) and $d_{x^2-y^2}$ (red color) orbitals, in which the former is more pronounced.

\begin{figure}[t]
\begin{center}
\includegraphics[width=1.0\columnwidth]{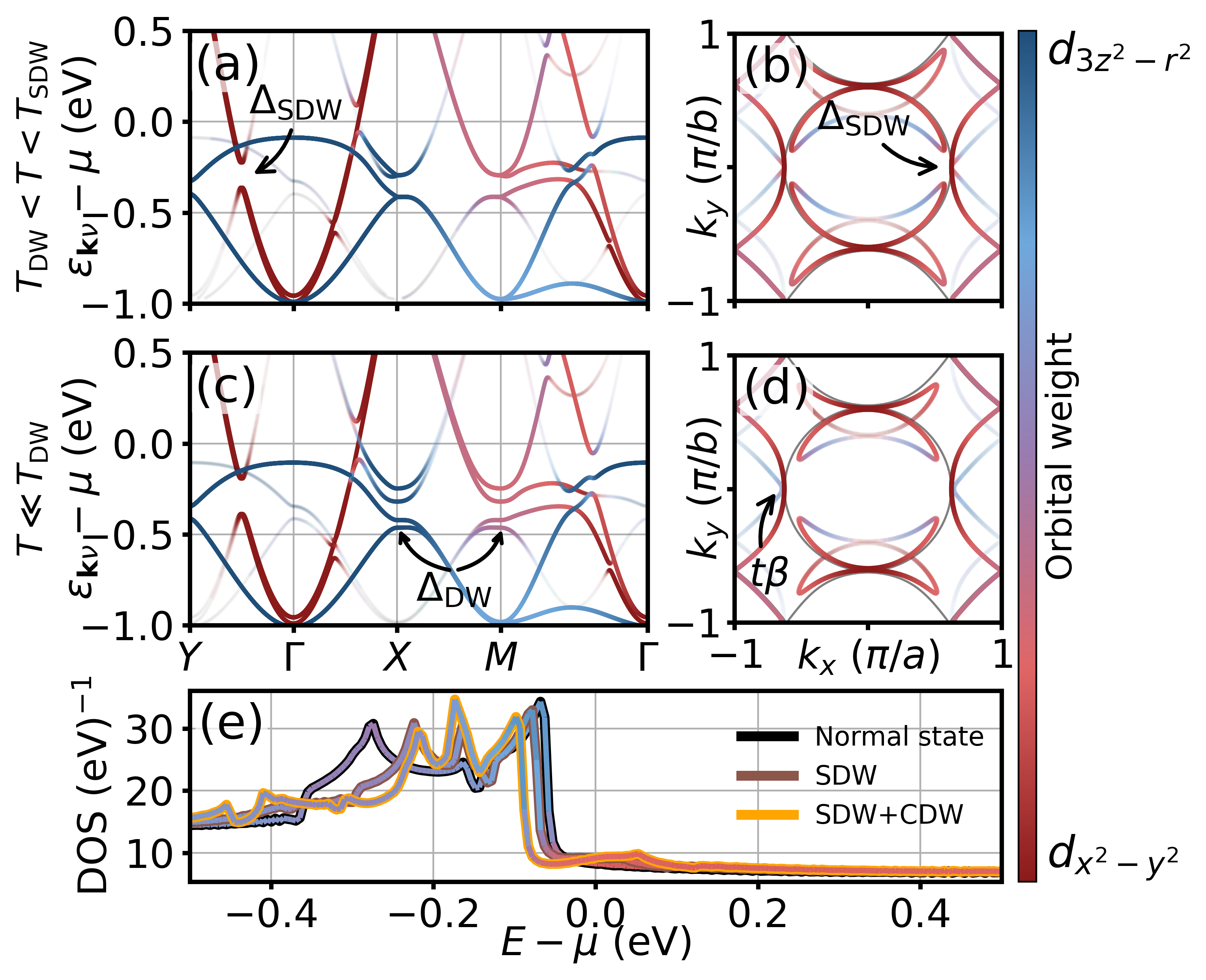}
\caption{ Calculated electronic band structure along high-symmetry directions (a,c), Fermi surfaces (b,d), and density of states (e) in the SDW and SDW/CDW ordered states.
Colors denote the $d_{3z^2-r^2}$ (blue) and $d_{x^2-y^2}$ (red) spectral weights and opacity quantifies the total spectral weight at each band and momentum tuple.
In (b,d), the SDW gap features are indicated, and thin gray lines show the normal-state band structure for comparison.
In panel (d), the folded $\beta$ pocket, dubbed $t\beta$ is indicated.
Self-consistent mean-field parameters $M_{\mu},\; \phi_{\mu}$, and $\mu$ for temperatures $T/T_{\text{SDW}}\approx0.78$ and $T/T_{\text{SDW}}\approx0.02$ are used in panels (a,b) and (c,d), respectively.
In (c), the density of states is shown as a function of energy for the normal (black contour), SDW (brown color), and SDW+CDW (orange contour) states.
Blue and red colors combine the integrated orbital contributions as in the band structure.
}
\label{fig:bands_ordered}
\end{center}
\end{figure}

At the characteristic temperature $T_{\text{DW}}$, a local symmetry is broken in the spin density: The in-plane Ni$_1$ and Ni$_2$ sites are no longer equivalent.
This is seen by a splitting in the dark (Ni$_1$) and light (Ni$_2$) colors in Fig.~\ref{fig:order_parameters}(a), which is pronounced as the purple and green arrows in Fig.~\ref{fig:orders}(b,d) acquire very different norms.
For same-spin sites, as e.g. dark triangles and light circles in Fig.~\ref{fig:order_parameters}(a), there is no preferential site to have suppressed or enhanced spin density below $T_{\text{DW}}$, therefore,  both solutions are degenerate in free energy.
This manifests as a random exchange of e.g. dark triangle and light circle indices in the temperature-dependent spin density inside the CDW phase. The local moment splitting is accompanied by a charge modulation that breaks in-plane site equivalency inside the unit cell.
The deviation from the average charge density per orbital is shown in the spheres of Fig.~\ref{fig:orders}, in which the Ni$_1$ (green) and Ni$_2$ (purple) feature opposite signs.
The charge-density splitting breaks in-plane site equivalency of the Ni$_1$A (Ni$_1$B) and Ni$_2$A (Ni$_2$B) sites for both $d_{3z^2-r^2}$ and $d_{x^2-y^2}$ orbitals at $T_{\text{DW}}$ [Fig. \ref{fig:order_parameters}(b)].
The magnetic channel features an order parameter of up to $\sim180$ meV, in striking contrast to the $\sim16$ meV charge order parameter, revealing a difference of one order of magnitude.
Still, the charge modulation coexists with a redistribution of spin density from the Ni$_2$ to the Ni$_1$ in-plane site, generating low-spin sites, as shown in Fig. \ref{fig:orders}(d).
This state is sometimes dubbed spin-charge stripe in the literature.
As shown in Fig. \ref{fig:orders}(c), the SDW+CDW state has as characteristic signature a kink at $T_{\text{DW}}$ in the temperature evolution of the magnetic moments per site.

Inside the magnetic phase, the folded bands become prominent and participate on the fermiology.
Fig.~\ref{fig:bands_ordered}(a) and (c) show the band structure in the the double-stripe (a) and charge-ordered low-spin-stripe (c) phases, where the effect of the $\Delta_{\text{SDW}}$ and $\Delta_{\text{DW}}$ gaps is highlighted.
The CDW plays the main role of splitting the band degeneracies present in the $\boldsymbol{X}$ and $\boldsymbol{M}$ high-symmetry points.
In the Fermi surface [panels (b) and (d)], the dominance of the $d_{x^2-y^2}$ orbital is clear.
This is a direct consequence of the larger magnetic moments at the $d_{3z^2-r^2}$ orbital, as discussed in Fig.~\ref{fig:order_parameters}(a).
However, panel (d) shows that the CDW enhances the $d_{3z^2-r^2}$ contribution to the Fermi surface, which panel (c) highlights that comes from a folded band from the $\boldsymbol{M}=(\pi,\pi)$ to the $\boldsymbol{X}$ high-symmetry point.
In fact, folded bands that loose spectral weight as a function of momentum and energy (frequently dubbed shadow bands) are more prominent around the $\boldsymbol{X}$ and $\boldsymbol{M}$ high-symmetry points.
Finally, Figs.~\ref{fig:bands_ordered}(b) and (d) indicate the folded $\beta$ pocket, dubbed $t\beta$ pocket is a direct consequence of the magnetic phase \cite{AuYeung2025UniversalNickelates}.
This folded pocket has mainly $d_{3z^2-r^2}$ orbital character.
The density of states is shown in Fig.~\ref{fig:bands_ordered}(e) for three temperature ranges: The normal state, the SDW, and the SDW+CDW state.
In the normal state (black contour), the $\gamma$ pocket dominates the density of states around the Fermi level and induces dominant $d_{3z^2-r^2}$ orbital contribution (blue color) in this range, whereas above the Fermi level the $d_{x^2-y^2}$ orbital (red color) dominates.
At $\sim0.3$ eV, there is another DOS peak associated with the doubly-degenerate bands at $\boldsymbol{X}$ and $\boldsymbol{M}$.
Lowering temperature (brown and orange contours) shifts the $d_{3z^2-r^2}$ bands away from the Fermi level, keeping the $d_{x^2-y^2}$ character at the Fermi energy, but also introduces a splitting of the $\sim0.3$ eV peak to a double-peak structure in the DOS.
This large energy splitting is associated with the spin-density wave gap, whereas an energy splitting caused by the charge-density wave gap is seen around $\sim-0.45$ eV.

\begin{figure}[t]
\begin{center}
\includegraphics[width=1.0\columnwidth]{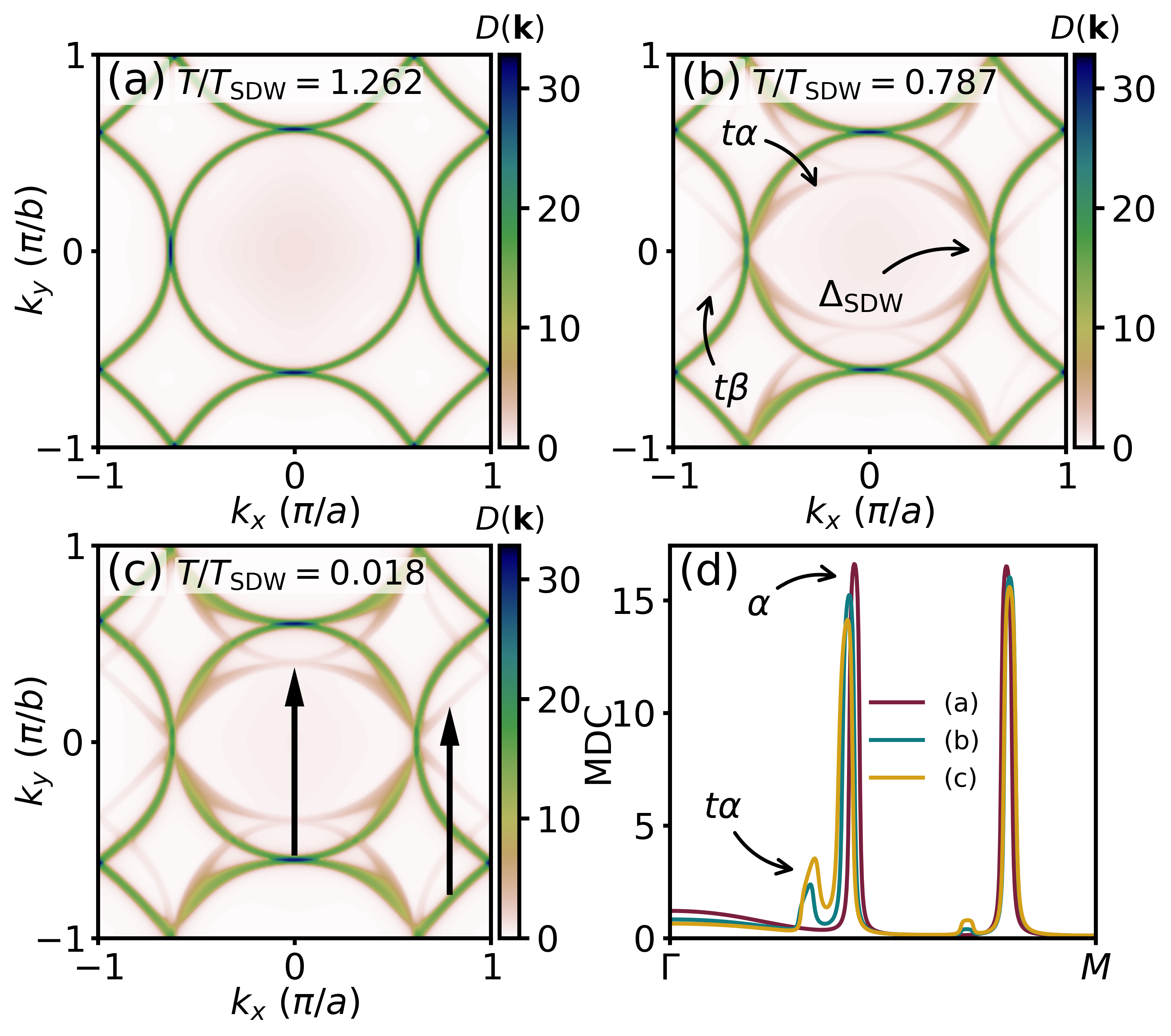}
\caption{ {Narrow-energy, temperature weighted spectral function $D(\boldsymbol{k})$ [Eq.~\eqref{eq:arpes}].
The first Brillouin zone is shown for temperatures $T>T_{\text{SDW}}$ (a), $T_{\text{SDW}}>T>T_{\text{DW}}$ (b), and $T_{\text{DW}}\gg T$ (c).
Panel (d) shows momentum-distribution curves (MDC) along the $\boldsymbol{\Gamma}-\boldsymbol{M}$ high-symmetry direction.}
}
\label{fig:arpes}
\end{center}
\end{figure}

{We further analyze the ARPES signatures of the density-wave states shown in Fig.~\ref{fig:bands_ordered}. ARPES probes the momentum- and energy-resolved spectral function, with the measured intensity additionally weighted by the temperature-dependent Fermi–Dirac distribution \cite{Sobota2021ARPES}. To emulate the experimental measurements, we integrate the Fermi–Dirac-weighted spectral function over a narrow energy window $[-\omega_0,\omega_0]$ around the Fermi level, as follows}
\begin{equation}
    D(\boldsymbol{k})=\int_{-\omega_0}^{\omega_0}d\omega A(\boldsymbol{k},\omega)f(\omega).
\label{eq:arpes}
\end{equation}
{
In Fig.~\ref{fig:arpes}, we show color maps of $D(\boldsymbol{k})$, calculated for an energy window centered at the Fermi level with $\omega_0=40$ meV. Panel (a) displays the normal-state Fermi-surface intensity map. In the density-wave states shown in panels (b) and (c), the overall Fermi-surface reconstruction remains relatively weak. Nevertheless, the spectral intensity along the $\boldsymbol{\Gamma}-\boldsymbol{Y}$ direction is higher than that along the $\boldsymbol{\Gamma}-\boldsymbol{X}$ direction. This anisotropy originates from the opening of the SDW gap near the points at which the $\alpha$ and $\beta$ pockets intersect.
Upon entering the CDW phase, no additional Fermi-surface features are clearly resolved. We nevertheless identify a translated (reconstructured) pocket inside the original $\alpha$ pocket, which we denote $t\alpha$, in analogy with the translated $t\beta$ pocket. The nesting vectors $\boldsymbol{Q}_Y$ responsible for generating the $t\alpha$ and $t\beta$ pockets are indicated in panel (c).
Figure~\ref{fig:arpes}(d) shows momentum-distribution curves (MDCs) along the high-symmetry direction $\boldsymbol{\Gamma}-\boldsymbol{M}$, obtained from cuts through the $D(\boldsymbol{k})$ maps presented in panels (a)–(c). As the temperature is lowered, the spectral intensity associated with the original $\alpha$ pocket is suppressed, while an additional peak at lower momentum, associated with the $t\alpha$ pocket, develops. For a direct comparison with experimental ARPES results, one should bear in mind that the measured spectral weight is additionally influenced by orbital- and photon-energy-dependent photoemission matrix elements, as well as by possible averaging over different orthorhombic domains \cite{AuYeung2025UniversalNickelates,li2026three}.}

\begin{figure}[t]
\begin{center}
\includegraphics[width=1.0\columnwidth]{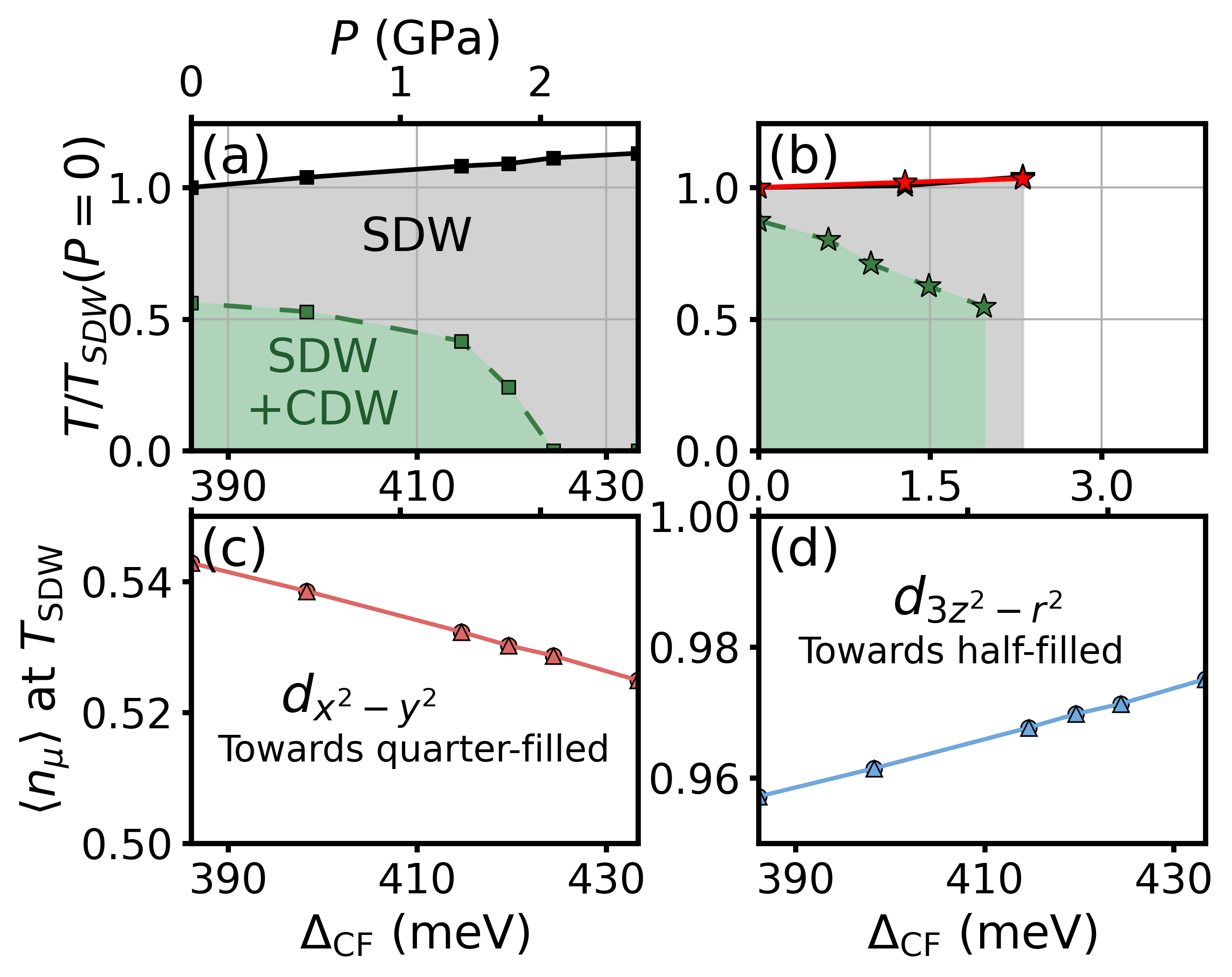}
\caption{ Calculated evolution of the $\boldsymbol{Q}_{Y}=(0,\pi)$ SDW order and the corresponding intra-unit cell charge order as a fucntion of the crystal field splitting  (a) and experimental (b) phase diagrams.
Experimental data is reproduced from \cite{Khasanov_PressureEnhancedSplittingDensityWaveTransitions_2025}.
$\Delta_{\text{CF}}$ denotes crystal-field splitting, used to map the theoretical results to the pressure axis.
The temperature axis is normalized by the ambient-pressure $T_{\text{SDW}}$ critical temperature.
{Orbital-resolved occupations for the $d_{x^2-y^2}$ (c) and $d_{3z^2-r^2}$ (d) calculated at the SDW onset temperature as a function of crystal-field splitting (bottom horizontal axis) and pressure (top horizontal axis).}
}
\label{fig:intro}
\end{center}
\end{figure}

We investigate the pressure evolution of the SDW and SDW+CDW states by tuning the crystal-field splitting of the ambient-pressure model.
By using the crystal-structure experimental data from Ref. \cite{Xu2026}, we performed crystal-field splitting DFT calculations for several pressure structures.
Details can be found in the Appendix \ref{appendix:computational}.
We conclude that crystal-field splitting monotonically increases with pressure evolution.
Although the band width and hopping integrals should also change with pressure{, these changes are small for lower pressures, because the lattice constants slowly change with this parameter such that} we argue that the difference in orbital content induced by crystal-field splitting is the main effect of pressure {in the low-pressure limit} \cite{Zhang2024StructuralNickelate}.
We qualitatively compare our results with the experimental phase diagram in Fig.~\ref{fig:intro}(a) and (b), respectively.
In fact, as a function of crystal-field splitting, the SDW critical temperature is enhanced and the CDW critical temperature suppressed, which qualitatively agrees with previous experimental data \cite{Khasanov_PressureEnhancedSplittingDensityWaveTransitions_2025}.
{
On the basis of this observation, we propose the main effects of pressure is the occupation of the $d_{x^2-y^2}$ and $d_{3z^2-r^2}$,
as shown in Fig.~\ref{fig:intro}(c) and (d), respectively, where we fix temperature at the onset of order $T_{\text{SDW}}$.
We emphasize that the orbital occupations are calculated self-consistently for finite $U$ and $J_H$ for the given crystal-field splitting.
The occupation of the $d_{x^2-y^2}$ orbital is suppressed towards quarter filling, while the $d_{3z^2-r^2}$-orbital occuptaion is enhanced towards half filling.
}

\section{ Discussions and conclusions }
\label{sec:discussions}
Our calculations reveal a metallic SDW/CDW state that is consistent with the extensive body of transport and spectroscopic measurements reported both inside and outside the ordered phases of La$_3$Ni$_2$O$_7$ \cite{Zhang1994LayeredNickelate,Liu2023ChargeSpinNickelate,Ling2000NeutronDiffractionNickelate,Chen2024ElectronicMagneticNickelate,Kakoi_MultibandMetallicGroundStateNickelates_2024,Fan2024TunnelingNickelate,Chen2025MultibandNickelate}. 
At ambient pressure, the calculated magnetic order parameter, $M_\mu\sim180$ meV, is also comparable to the experimentally inferred SDW gap, $2\Delta\sim65$--100 meV \cite{Liu2024ElectronicCorrelationsNickelate,Meng2024DensityWaveNickelate,Fan2024TunnelingNickelate,Wu2025UltrafastDensityWaves,Zhu2026HighEnergyExcitationsNickelate,Kakoi_MultibandMetallicGroundStateNickelates_2024}.
A key benchmark for the theory is the sequence of density-wave transitions. 
Multiple experimental probes consistently report transition temperatures of approximately $T_{\text{SDW}}\sim150$ K and $T_{\text{DW}}\sim130$ K, corresponding to $T_{\text{DW}}/T_{\text{SDW}}\sim0.87$ \cite{Puphal2026NickelateReview,Khasanov2026IsotopeNickelate,Liu2026DensityWavePhases,Wu2001MagneticSusceptibilityNickelate,Hosoya2008PressureNickelate,Zhang2024HTSCNickelate,Ling2000NeutronDiffractionNickelate,Liu2024ElectronicCorrelationsNickelate,Liu2023ChargeSpinNickelate,Chen2024SpinDensityWaveNickelate,Khasanov_PressureEnhancedSplittingDensityWaveTransitions_2025,Wu2025UltrafastDensityWaves,Meng2024DensityWaveNickelate,Wang2024PressureSuperconductivityNickelate,Luo_MicroscopicEvidenceChargeSpinDensityWaves_2025,Kakoi_MultibandMetallicGroundStateNickelates_2024}. 
Although our calculations yield a somewhat smaller ratio of $\sim0.6$, they capture several defining features of the ordered state. 
Experimentally, the SDW has been identified as a double-stripe order \cite{Plokhikh2025SpinDensityNickelate,Gupta2025SpinStripeNickelate,Zhou2025PressureSDWNickelate}, with evidence for spin-poor sites within the magnetic unit cell \cite{Yashima2025SpinStripeNickelate,Khasanov_PressureEnhancedSplittingDensityWaveTransitions_2025,Zhou2026SpinFluctuationsNickelates,Chen2026NatureMagnetismNickelate}.
Raman measurements further indicate that the SDW gap is strongly anisotropic and momentum dependent \cite{Shu2026MomentumSelectiveSDW,Gim2026SpectroscopicNickelate}.
Consistent with these observations, our results support a commensurate double-stripe SDW with ordering vector $\boldsymbol{Q}_Y$ above $T_{\text{DW}}$, characterized by a highly momentum-dependent Fermi-surface gap and orbital-selective magnetic order. 
Upon cooling below $T_{\text{DW}}$, a charge-density wave emerges together with low-spin sites. 
A direct consequence of this transition is a kink in the temperature dependence of the local magnetic moments. 
{The observation of the kink in experiment should be considered as a direct proof of the coupling of the intertwined charge and spin orders in La$_3$Ni$_2$O$_7$ and is one of the predictions of the present theory.}  

{It is interesting to note that $\mu$SR and $^{139}$La NQR measurements have been already argued to be better fit by a double-stripe-spin-spinless order  \cite{Chen2024SpinDensityWaves,Khasanov_PressureEnhancedSplittingDensityWaveTransitions_2025,Khasanov2026PressureIsotopeLa4Ni3O10,Khasanov2026IsotopeNickelate,Luo_MicroscopicEvidenceChargeSpinDensityWaves_2025} consistent with our calculations {for the spin-modulated double stripe phase if one takes into account that experiment cannot exclude finite yet small magnetic moments on the spinless Ni-sites}. 
The only caveat is that it was conjectured that such a spin spinless stripe order appears right at $T_{\text{SDW}}$, as it yields the experimental observed two types of internal magnetic fields, stemming from spin-full and spin-poor regions \cite{Khasanov_PressureEnhancedSplittingDensityWaveTransitions_2025}. 
Within our theory the double spin stripe order develops first at $T_{\text{SDW}}$ yet the double-stripe-spin-spinless order follows at lower temperature $T_{\text{DW}}$, which yields additional kink in the temperature dependence of the internal  magnetic fields. In principle, the appearance of the double stripe spin-full spin-poor states already at $T_{\text{SDW}}$ cannot be excluded within our theory yet it would likely impose the first order phase transition due to linear coupling of the SDW and DW orders, which is not confirmed experimentally. 
One of the reasons for the discrepancy could be that  our starting model is based on the $Amam$ crystal space group, in which the four Ni sites in the unit cell are related by symmetry operations. The electronic charge order then breaks such symmetry at a temperature $T_{\text{DW}}<T_{\text{SDW}}$, which is, therefore, unrelated to the spinless sites argued by $\mu$SR and NQR theoretical analyses. 
At the same time, 
recent synchrotron X-ray diffraction measurements have reported a high-temperature (above $T_{\text{SDW}}$) structural charge order to the space group {$Am2m$} \cite{Misawa2026,Reiss2026,Liu2026DensityWavePhases,Sasaki2026StructuralPhaseDiagram}, which breaks the site-equivalence symmetry and would then allow our model to find solutions for the SDW at $T_{\text{SDW}}$ with spinless sites.
Further investigations are needed to confirm this conjecture.}

The orbital structure of the SDW provides a possible interpretation of several spectroscopic observations. At low temperatures and ambient pressure, the calculated $d_{x^2-y^2}$ and $d_{3z^2-r^2}$ magnetic order parameters differ only slightly, with a ratio of approximately 0.95. We suggest that this orbital dependence is related to the two-component SDW gap observed in time-resolved optical spectroscopy, where gaps of $\Delta_1\sim67$ meV and $\Delta_2\sim54$ meV were reported, corresponding to a ratio of $\sim0.81$ \cite{Zhu2026HighEnergyExcitationsNickelate}. Tunneling experiments have also identified two low-energy density-of-states peaks located at negative and positive bias \cite{Fan2024TunnelingNickelate}. The negative-bias feature has been associated with the SDW, whereas the positive-bias peak was attributed to a correlated $d_{x^2-y^2}$ background. In our calculations, the SDW and CDW primarily modify the occupied-state peak, while no analogous high-energy feature appears in the density of states. The transfer of $d_{x^2-y^2}$ spectral weight toward the Fermi level likely originates from correlation effects beyond the present mean-field description, but is captured, for example, by DMFT calculations \cite{Leonov2025SpinChargeStripesNickelate,Liu2025OrbitalSelectivityNickelate,Wang2024ElectronicMagneticNickelate}.

Having established the magnetic properties of the ordered phase, we now turn to the nature of the CDW transition. Local probes provide compelling evidence that this transition has a charge character in La$_3$Ni$_2$O$_7$ \cite{Liu2023ChargeSpinNickelate,Kakoi_MultibandMetallicGroundStateNickelates_2024,Luo_MicroscopicEvidenceChargeSpinDensityWaves_2025,Khasanov2026IsotopeNickelate}. In particular, recent $^{139}$La NQR measurements point to a commensurate unidirectional CDW \cite{Luo_MicroscopicEvidenceChargeSpinDensityWaves_2025}. Our results support such an interpretation and suggest that the charge modulation is electronic in origin, producing a stripe-like pattern within the unit cell. The resulting CDW is strongly orbital dependent, with the dominant contribution arising from the $d_{x^2-y^2}$ orbital, in contrast to the magnetic order parameter. Microscopically, the instability originates from an imbalance between the Ni$_1$ and Ni$_2$ local moments, generating clear energy splittings in the density of states. Since the characteristic energy scale of the CDW is only $\sim16$ meV, its signatures are expected to be most pronounced in low-energy probes such as transport measurements.

An important consequence of the proposed CDW is the breaking of {sublattice symmetry} within the crystallographic unit cell, rendering the Ni$_1$ and Ni$_2$ sites inequivalent, albeit by a small amount. This observation may help explain the pronounced sensitivity of the CDW transition to the oxygen environment. Isotope-substitution experiments have shown that the CDW in bilayer nickelates is strongly affected by oxygen-related structural changes, in contrast to the behavior observed in trilayer compounds \cite{Khasanov2026IsotopeNickelate,Khasanov2026PressureIsotopeLa4Ni3O10}. Because the instability proposed here is low energy and relies on the inequivalence of in-plane sites, it should be particularly susceptible to structural modifications that restore site symmetry. Pressure provides precisely such a mechanism by driving the system toward a tetragonal crystal structure. We therefore conjecture that strain engineering aimed at equalizing the in-plane lattice constants at ambient pressure should similarly suppress the CDW transition by restoring the symmetry relation between Ni$_1$ and Ni$_2$.

The proposed SDW/CDW ordered state also accounts for several key ARPES observations. Experiments consistently place the $\gamma$ pocket approximately 100 meV below the Fermi level and report only modest Fermi-surface reconstruction across the SDW transition \cite{Yang_OrbitalDependentElectronCorrelationLa3Ni2O7_2024,Li2024PseudogapNickelate,AuYeung2025UniversalNickelates}. 
{Although the $Amam$ space group breaks $C_4$ symmetry in the normal state already, this effect is very small ($a/b\approx98\%$) an no changes in the Fermi surface are visible in the normal state.
Inside the density wave states, however, we see $(i)$ the modest reconstruction of the Fermi surface, $(ii)$ higher spectral intensity along the $\boldsymbol{\Gamma}-\boldsymbol{Y}$ direction than along the $\boldsymbol{\Gamma}-\boldsymbol{X}$ path, and $(iii)$ that the CDW has little impact in the Fermi surface.
The SDW reconstruction yields the $t\beta$-pocket, which might be often hidden from the more intense signals of the $\alpha$ and $\beta$ pockets, which presents a reasonable explanation for the lack of evidence of SDW-driven Fermi surface reconstructions in some ARPES reports \cite{Yang_OrbitalDependentElectronCorrelationLa3Ni2O7_2024,Li2024PseudogapNickelate}.
This low intensity is a consequence of a strong energy dependence of $t\beta$, which has been reported in recent experiments \cite{AuYeung2025UniversalNickelates}.
We also see another translated pocket inside the $\alpha$ pocket, as recently reported in \cite{AuYeung2025UniversalNickelates}, which we also dub $t\alpha$.
From our model, $t\alpha$  and $t\beta$ arise from the same $\boldsymbol{Q}_Y$ band folding, which taken together  with Hartree shifts do not require a second nesting vector as argued in \cite{AuYeung2025UniversalNickelates}.
Our calculations predict a decreasing intensity of the $\alpha$ pocket ARPES signal, which has also been reported in \cite{AuYeung2025UniversalNickelates}.}

Beyond its spectroscopic signatures, the SDW has important implications for rotational symmetry breaking. 
Resonant soft x-ray scattering, polarimetry, and polarized ultrafast spectroscopy have all suggested that the density-wave state of La$_3$Ni$_2$O$_7$ possesses an electronic nematic character \cite{Wu2025UltrafastDensityWaves,Gupta2025SpinStripeNickelate,Wu2026ElectronicNematicityNickelate}. 
Importantly, the observed symmetry breaking cannot be attributed solely to the weak orthorhombicity of the lattice {($a/b\sim98\%$ \cite{Xu2026,Plokhikh2025SpinDensityNickelate})}, which has only a minor effect on the normal-state electronic structure. {Instead, nematicity emerges because the system further deviates from $C_4$ rotational symmetry below $T_{\text{SDW}}$.}
This behavior arises naturally within our framework. 
The single-$\boldsymbol{Q}_Y$ SDW {strongly deviated from fourfold rotational symmetry enlarging the crystallographic unit cell into a magnetic unit cell with pronounced anisotropy.} 
Consequently, the electronic symmetry breaking is strongly amplified inside the ordered phase.

Finally, our results provide a natural explanation for the contrasting pressure evolution of the SDW and CDW. 
Experiments show that the CDW is progressively suppressed under pressure, whereas the SDW remains comparatively robust \cite{Khasanov_PressureEnhancedSplittingDensityWaveTransitions_2025,Zhou2025PressureSDWNickelate,Khasanov2026IsotopeNickelate,Liu2026DensityWavePhases}. 
In our theory, the unidirectional CDW breaks the equivalence between in-plane Ni sites and induces band splittings near the $\boldsymbol{X}$ and $\boldsymbol{M}$ points, where the normal and SDW states exhibit twofold degeneracies. 
By analyzing the evolution of the ordered state as a function of crystal-field splitting, we find that pressure gradually restores the in-plane site symmetry and thereby suppresses the CDW. 
This interpretation is consistent with the structural transition from an orthorhombic ($b>a$) to a tetragonal ($a=b$) lattice under pressure.
{We predict that the $d_{3z^2-r^2}$ orbital is more relevant for the magnetic order, while the $d_{x^2-y^2}$ orbital plays a more important role for the charge order [Fig.~\ref{fig:order_parameters}].
As a consequence, the $d_{3z^2-r^2}$ orbital contribution, enhanced by pressure towards half-filling, contributes to enhance $T_{\text{SDW}}$ and the suppression of $d_{x^2-y^2}$ orbital contribution as pressure evolves leads to a suppression of $T_{\text{DW}}$.}
The accompanying redistribution of spin density is likewise weakened as the CDW disappears. 
{The small enhancement of $T_{\text{SDW}}$ with pressure can be understood having in mind the cooperation between $d_{3z^2-r^2}$ and $d_{x^2-y^2}$ contributions to the magnetic order, in which $d_{3z^2-r^2}$ is slightly more important such that the suppression of the $d_{x^2-y^2}$ contribution is slightly overcome by the enhanced $d_{3z^2-r^2}$ contribution.}
Future spectroscopic probes sensitive to band splittings, together with local probes of magnetic moments, should therefore provide stringent tests of the proposed pressure phase diagram.

In summary, we propose that the low-pressure phase diagram of La$_3$Ni$_2$O$_7$ is governed by a double-stripe SDW that develops below $T_{\text{SDW}}$, followed at lower temperatures by the onset of a low-energy unidirectional intra-unit-cell CDW at $T_{\text{DW}}$. 
The CDW is accompanied by a redistribution of spin density that generates low-spin sites within the magnetic unit cell. 
Furthermore, our analysis of the pressure dependence of the ordered phases identifies the crystal-field splitting as a key parameter controlling the ratio $T_{\text{DW}}/T_{\text{SDW}}$ and the eventual suppression of the CDW in the high-pressure regime.

\begin{acknowledgments}
We thank Takasada Shibauchi, Thomas J. Hicken, {and Andrea Damascelli} for fruitful discussions.
L.B.B. acknowledges financial support from the grants 2023/14902-8 and 2025/17852-7, São Paulo Research Foundation (FAPESP).
L.G.G.V.D.S. acknowledges financial support by Brazilian agencies CNPq (Grant. No. 312622/2023-6), and FAPESP (Grant no. 2022/15453-0).
\end{acknowledgments}

\section{Data availability}
The data that support the findings of this paper are openly available on Zenodo \cite{ZenodoData_action}.

\appendix
\section{Computational details}
\label{appendix:computational}

In this appendix, we provide computational details, including details on the mean-field equation solver, convergence tests, and the DFT calculations performed for the present work.
Further analyses of parameter exploration and data can be found in the Python notebooks, available together with our code in the Zenodo repository [ref.].

We numerically implement the non-interacting model of Eq.~\eqref{eq:H0_sublattice} using the TRIQS library \cite{ParcolletTRIQSToolboxForResearch2015}.
This code is actively maintained and efficiently deals with real-space hopping parameters and intricate details of all sorts of lattice symmetries.

To solve mean-field equations, which are obtained by equating the jacobian of the free energy functional to zero that is, $\boldsymbol \nabla F = 0$, we minimize the multivariate grand potential using the Broyden–Fletcher–Goldfarb–Shanno algorithm \cite{Broyden1970DoubleRankMinimization,Fletcher1970VariableMetric,Goldfarb1970VariableMetric,Shanno1970ConditioningQuasiNewton} as implemented by the standard \textit{python} library \textit{scipy}.
The function to be minimized is the grand potential $\Omega=F+\mu n$, where the free energy $F$ is given by Eq.~\eqref{eq:FMF_Kanamori}.
As can be noted in Eq.~\eqref{eq:mu_kanamori} derived in Appendix \ref{appendix:action}, the equation for the chemical potential $\mu$ is a simpler one-dimensional problem.
We, therefore, solve the chemical potential constraint implicitly by the nearly-exact one-dimensional Brent’s method \cite{Brent1973} (\textit{brentq} function of \textit{scipy.optimize}).
We set $n=1.5$ per Ni layer \cite{Zhang2023ElectronicStructureNickelate}.
In our openly available code, an example of application is given for the $\boldsymbol{Q}=(\pi,\pi)$ square lattice Hubbard model, where the free energy and its jacobian vector field can be visualized in a two-dimensional map and the minima clearly spotted.

\begin{figure}[t]
\begin{center}
\includegraphics[width=1.0\columnwidth]{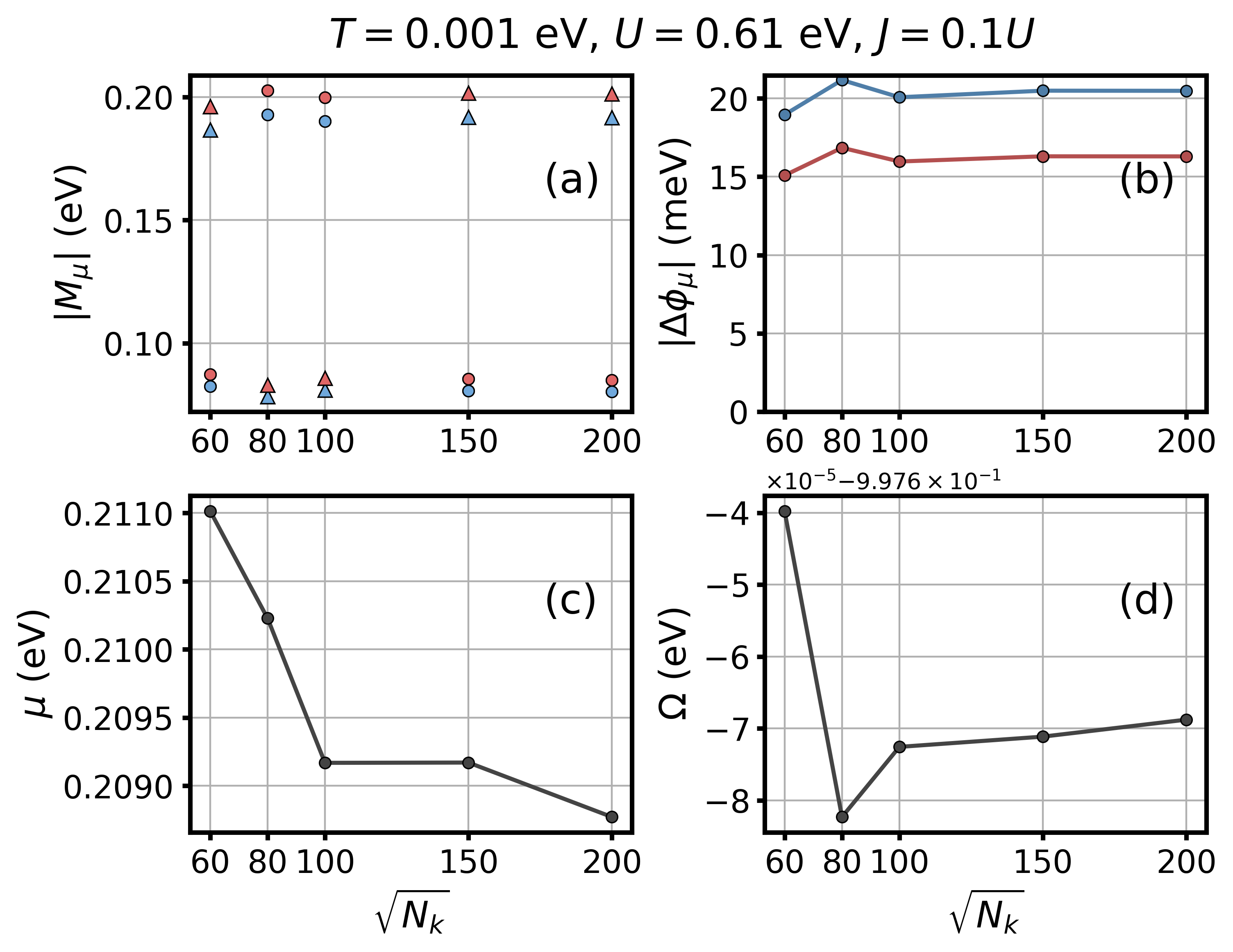}
\caption{ Convergence tests of the momentum grid for the magnetic (a) and charge (b) order parameters, chemical potential (c), and free energy (d).
In panels (a) and (b), the color code follows Fig.~\ref{fig:order_parameters}.
}
\label{fig:convergence}
\end{center}
\end{figure}

We carefully checked convergence of the relevant quantities for the present problem.
Fig.~\ref{fig:convergence} shows the magnetic [panel (a)] and charge [panel (b)] order parameters, chemical potential [panel (c)], and the grand potential [panel (d)] as a function of the number of momentum points in the two-dimensional grid $\sqrt{N_k}\times\sqrt{N_k}$.
We used $\sqrt{N_k}=100$ in all reported simulations.
The \textit{scipy} algorithm also has internal convergence criteria, to which all the reported simulations passed.
To ensure an unbiased minimization procedure, we ran 5 simulations with independent random initial guesses for each temperature point in the phase diagram and selected the solution with minimum grand potential energy.

The effective single-particle Hamiltonian $H_0$ of tight-binding form is obtained from DFT calculations and subsequent Wannier construction based on the maximally-localized procedure~\cite{marzari12}. The DFT computations are performed within the local-density approximation using a mixed-basis pseudopotential framework~\cite{louie1979,elsaesser90,lechermann02,mbpp_code}. This technique utilizes norm-conserving pseudopotentials, as well as plane waves and atomic-like localized functions to represent the crystal wave functions. The plane-wave cutoff was chosen $E_{\rm cut}=16$\,Ryd and localized functions are introduced for La$(5d)$, Ni$(3d)$, O$(2s)$ and O$(2p)$. For the primitive cells encircling 6 La, 4 Ni and 14 O sites, a $k$-point mesh of size $5\times 5\times 5$ is employed. The maximally-localized Wannier construction results in an effective $8\times 8$ Hamiltonian
for the $\{d_{z^2},d_{x^2-y^2}\}$ orbitals of the four symmetry-equivalent Ni atoms in the primitive cell. The crystal-field data in Table \ref{tab:cf} are derived from the respective onsite terms of the constructed Wannier Hamiltonians.

\begin{table}[t]
\caption{Evolution of crystal-field splitting $\Delta_{\text{CF}}=\phi_{d_{x^2}}-\phi_{d_{z^2}}$ with pressure computed by DFT using the experimental structure data from Ref.~\cite{Xu2026}. We mapped our results in the pressure axis using the function $P(\Delta_{\text{CF}})$ fitted from this data.}
\begin{ruledtabular}
\begin{tabular}{c|c|c|c|c|c}
Pressure (GPa)               & 0.00  & 1.19  & 2.53  & 3.88  & 5.94  \\ \hline
Crystal-field splitting (eV) & 0.386 & 0.412 & 0.430 & 0.433 & 0.435
\end{tabular}
\end{ruledtabular}
\label{tab:cf}
\end{table}

\section{Spin and charge orders in the action formalism}
\label{appendix:action}
In this appendix, we detail the spin and charge order parameters as probed for in our unrestricted Hartree-Fock calculations.
We consider the multiorbital Hamiltonian
\begin{align}
H = H_0 + H_{\mathrm{int}},
\end{align}
with the quadratic and quartic parts given by Eqs.~\eqref{eq:H0_sublattice} and \eqref{eq:Kanamori_nc_form}, respectively.
The interaction tensors in $H_{\mathrm{int}}$ [Eq.~\eqref{eq:Kanamori_nc_form}] are given by
\begin{align}
(U_s)^{\mu\eta}_{\alpha\beta}
=
\begin{cases}
U, & \mu=\eta=\alpha=\beta, \\
U', & \mu=\eta\neq\alpha=\beta, \\
J_H, & \mu=\alpha\neq\eta=\beta, \\
J_H', & \mu=\beta\neq\eta=\alpha, \\
0, & \text{otherwise}.
\end{cases}\label{eq:Us}
\end{align}
\begin{align}
(U_c)^{\mu\eta}_{\alpha\beta}
=
\begin{cases}
U, & \mu=\eta=\alpha=\beta, \\
- U' + 2J_H, & \mu=\eta\neq\alpha=\beta, \\
- J_H + 2U', & \mu=\alpha\neq\eta=\beta, \\
J_H', & \mu=\beta\neq\eta=\alpha, \\
0, & \text{otherwise},\label{eq:Uc}
\end{cases}
\end{align}
where it is implicit that the index combination must be constrained to be onsite.

Introducing Grassmann fields
$c_{\boldsymbol{k}\mu\sigma}(\tau)$ and
$c^{\dagger}_{\boldsymbol{k}\mu\sigma}(\tau)$,
the action for the Hamiltonian $H$ is
\begin{align}
S
=
\int_0^\beta d\tau
\Bigg[
\sum_{\boldsymbol{k}\mu\eta\sigma}
c^{\dagger}_{\boldsymbol{k}\mu\sigma}
(\partial_\tau \delta_{\mu\eta} + \xi^{\mu\eta}_{\boldsymbol{k}})
c_{\boldsymbol{k}\eta\sigma}
+
H_{\mathrm{int}}(\tau)
\Bigg].
\end{align}
Explicitly,
\begin{align}
S
=
\int_0^\beta d\tau
\sum_{\boldsymbol{k}\mu\eta\sigma}
c^{\dagger}_{\boldsymbol{k}\mu\sigma}
(\partial_\tau \delta_{\mu\eta} + \xi^{\mu\eta}_{\boldsymbol{k}})
c_{\boldsymbol{k}\eta\sigma}
+
S_{\mathrm{int}},
\end{align}
with
\begin{equation}
\begin{split}
    S_{\mathrm{int}}
=
\frac{1}{2N}
\int_0^\beta d\tau
\sum_{\boldsymbol{q}}
&\sum_{\mu\eta\alpha\beta}
\Big[
(U_c)^{\mu\eta}_{\alpha\beta}
\, n_{-\boldsymbol{q}\mu\eta}
\, n_{\boldsymbol{q}\alpha\beta}
\\
&-
(U_s)^{\mu\eta}_{\alpha\beta}
\, \boldsymbol{S}_{-\boldsymbol{q}\mu\eta}
\cdot
\boldsymbol{S}_{\boldsymbol{q}\alpha\beta}
\Big].
\end{split}
\end{equation}
This action is quartic in fermionic fields. 
We then introduce charge and spin auxiliary fields
$\phi_{\boldsymbol{q}\mu\eta}(\tau)$ and
$\boldsymbol{M}_{\boldsymbol{q}\mu\eta}(\tau)$.
{In analogy to a Gaussian quadrature rule,} the Hubbard-Stratonovich transformation reads
\begin{align*}
&
\frac{1}{2N}
(U_c)^{\mu\eta}_{\alpha\beta}
\, n_{-\boldsymbol{q}\mu\eta}
\, n_{\boldsymbol{q}\alpha\beta}
-
\frac{1}{2N}
(U_s)^{\mu\eta}_{\alpha\beta}
\, \boldsymbol{S}_{-\boldsymbol{q}\mu\eta}
\cdot
\boldsymbol{S}_{\boldsymbol{q}\alpha\beta}
\nonumber\\[6pt]
\longrightarrow
&
\frac{1}{N}
\phi_{\boldsymbol{q}\mu\eta}
\, n_{-\boldsymbol{q}\eta\mu}
-
\frac{1}{N}
\boldsymbol{M}_{\boldsymbol{q}\mu\eta}
\cdot
\boldsymbol{S}_{-\boldsymbol{q}\eta\mu}
\nonumber\\[6pt]
&
+
\frac{1}{2N}
\phi_{-\boldsymbol{q}\mu\eta}
(U_c^{-1})^{\mu\eta}_{\alpha\beta}
\phi_{\boldsymbol{q}\alpha\beta} \\
&+
\frac{1}{2N}
\boldsymbol{M}_{-\boldsymbol{q}\mu\eta}
\cdot
(U_s^{-1})^{\mu\eta}_{\alpha\beta}
\boldsymbol{M}_{\boldsymbol{q}\alpha\beta}.
\end{align*}
Then, the fermionic part becomes
\begin{equation}
\begin{split}
S_{\mathrm{ferm}}
&=
\int_0^\beta d\tau
\sum_{\boldsymbol{k} \boldsymbol{k}'\mu\eta\sigma\sigma'}
c^{\dagger}_{\boldsymbol{k}\mu\sigma}
\Big[
\partial_\tau \delta_{\boldsymbol{k}\boldsymbol{k}'}\delta_{\mu\eta}\delta_{\sigma\sigma'}\\
&+
\xi^{\mu\eta}_{\boldsymbol{k}} \delta_{\boldsymbol{k}\boldsymbol{k}'}
+
\phi_{(\boldsymbol{k}-\boldsymbol{k}')\mu\eta}
-
\boldsymbol{M}_{(\boldsymbol{k}-\boldsymbol{k}')\mu\eta}
\cdot
\boldsymbol{\sigma}_{\sigma\sigma'}
\Big]
c_{\boldsymbol{k}'\eta\sigma'}.
\end{split}
\end{equation}

Since the fermionic action is quadratic, we integrate out the fermions exactly performing a Gaussian Berezin integration such that the partition function becomes
\begin{align}
Z
=
\int
\mathcal{D}[\hat{\phi},\hat{\boldsymbol{M}}]
\, e^{-S_{\mathrm{eff}}[\hat{\phi},\hat{\boldsymbol{M}}]},\label{eq:Z}
\end{align}
with
\begin{equation}
\begin{split}
S_{\mathrm{eff}}
&=
\frac{1}{2N}
\int_0^\beta d\tau
\sum_{\boldsymbol{q}}
\Big[
\hat{\phi}_{-\boldsymbol{q}} \hat{U}_c^{-1} \hat{\phi}_{\boldsymbol{q}}
+
\hat{\boldsymbol{M}}_{-\boldsymbol{q}}\cdot \hat{U}_s^{-1} \hat{\boldsymbol{M}}_{\boldsymbol{q}}
\Big] \\
&-
\mathrm{Tr}\ln
\Big[
\hat{G}^{-1}(\hat{\phi},\hat{\boldsymbol{M}})
\Big],
\label{eq:Seff_Kanamori}
\end{split}
\end{equation}
where the orbital indices have been dropped to give room for a matrix notation and the inverse Green's function reads
\begin{align}
\hat{G}^{-1}_{\boldsymbol{k} \boldsymbol{k}'}
=
\hat{\partial}_\tau \delta_{\boldsymbol{k}\boldsymbol{k}'}
+
\hat{H}_{\boldsymbol{k}\boldsymbol{k}'},
\label{eq:GF_Kanamori}
\end{align}
where
\begin{equation}
    \hat{H}_{\boldsymbol{k}\boldsymbol{k}'} = \hat{\xi}_{\boldsymbol{k}}\delta_{\boldsymbol{k}\boldsymbol{k}'} + \hat{\phi}_{\boldsymbol{k}-\boldsymbol{k}'}
-
\hat{\boldsymbol{M}}_{\boldsymbol{k}-\boldsymbol{k}'}
\cdot
\boldsymbol{\sigma}
\label{eq:action_Hamiltonian}
\end{equation}
is the general Hubbard-Stratonovich Hamiltonian.

We now perform the saddle-point (mean-field) approximation of the effective action in Eq.~\eqref{eq:Seff_Kanamori}.
The saddle point is defined by
\begin{align}
\frac{\delta S_{\mathrm{eff}}}{\delta \phi_{\boldsymbol{q}\mu\eta}} = 0,
\qquad
\frac{\delta S_{\mathrm{eff}}}{\delta \mathbf{M}_{\boldsymbol{q}\mu\eta}} = 0.
\end{align}
Using Eq.~\eqref{eq:Seff_Kanamori}, we obtain
\begin{align}
-\sum_{\alpha\beta}
(U_c^{-1})^{\mu\eta}_{\alpha\beta}
\phi_{\boldsymbol{q}\alpha\beta}
&=
\langle n_{\boldsymbol{q}\eta\mu} \rangle,
\\[6pt]
\sum_{\alpha\beta}
(U_s^{-1})^{\mu\eta}_{\alpha\beta}
\boldsymbol{M}_{\boldsymbol{q}\alpha\beta}
&=
\langle \boldsymbol{S}_{\boldsymbol{q}\eta\mu} \rangle.
\end{align}
Thus, the auxiliary fields acquire the physical interpretation
\begin{align}
\phi_{\boldsymbol{q}\mu\eta}
&=
-\sum_{\alpha\beta}
(U_c)^{\mu\eta}_{\alpha\beta}
\langle n_{\boldsymbol{q}\alpha\beta} \rangle,
\\[6pt]
\boldsymbol{M}_{\boldsymbol{q}\mu\eta}
&=
\sum_{\alpha\beta}
(U_s)^{\mu\eta}_{\alpha\beta}
\langle \boldsymbol{S}_{\boldsymbol{q}\alpha\beta} \rangle.
\end{align}
Notice that $\phi_{0\mu\eta}$ is the known Hartree correction.
$\boldsymbol{M}_{\boldsymbol{q}\mu\eta}$ and $\phi_{\boldsymbol{q}\mu\eta}$ describe orbital-resolved magnetic order and charge density modulations at wavevector $\boldsymbol{q}$, respectively.
We remark that $\boldsymbol{q}$ is not the same for the charge and magnetic order parameters.
Here, the magnetic wave vector is $\boldsymbol{q}=\boldsymbol{Q}$ and the charge is $\boldsymbol{q}=2\boldsymbol{Q}=0$.
We also restrict to static, orbital-diagonal saddle-point fields such that
\begin{align}
\phi_{\boldsymbol{q}\mu\eta}(\tau) &\rightarrow \phi_{\mu}\delta(\tau)\delta_{\mu\eta}\delta_{\boldsymbol{q}0}\\
\boldsymbol{M}_{\boldsymbol{q}\mu\eta}(\tau) &\rightarrow \boldsymbol{M}_{\mu}\delta(\tau)\delta_{\mu\eta}\delta_{\boldsymbol{q}\boldsymbol{Q}}.
\end{align}
We then define the Nambu operators
\begin{align}
\Psi_{\boldsymbol{k}\sigma}
=
\left(
c_{\boldsymbol{k}\mu\sigma},
c_{(\boldsymbol{k}+\boldsymbol{Q})\mu\sigma}
\right)^{T}
\label{eq:spinor_kanamori}
\end{align}
to write the Hamiltonian of Eq.~\eqref{eq:action_Hamiltonian} as the main text form of Eq.~\eqref{eq:HMF}.
In doing so, we are doubling the number of sites and orbitals in the Hamiltonian, which is the effect of unit cell folding, implying that the Brillouin zone halves.
To avoid double counting momenta, a reduced Brillouin zone (RBZ) is defined with half of the area of the orthorhombic Brillouin zone.
We implement the RBZ by considering only half of the $k_x$-direction momenta.

The trace-log term in Eq.~\eqref{eq:Seff_Kanamori} becomes
\begin{align}
\mathrm{Tr}\ln G^{-1}
=
\sum_{\boldsymbol{k}\in RBZ}
\sum_{\nu\sigma}
\sum_{\omega_m}
\ln\left(-i\omega_m + \epsilon_{\boldsymbol{k}\nu}-\mu\right).
\end{align} 
Performing the Matsubara summation for the fermionic frequencies $i\omega_m$,
\begin{align}
\sum_{\omega_m}
\ln\left(-i\omega_m + E\right)
=
\ln\left(1+e^{-\beta E}\right).
\end{align}
With this, and noticing that, in the saddle-point solution, Eq.~\eqref{eq:Z} implies $S_{\text{eff}}=\beta F$, Eq.~\eqref{eq:FMF_Kanamori} is obtained.

We now derive the jacobian of the mean-field free energy obtained in the previous subsection, within the collinear and intraorbital approximation.
The fermionic Hamiltonian for each spin $\sigma=\pm1$ is diagonalized in the reduced Brillouin zone (RBZ) using the spinor in Eq.~\eqref{eq:spinor_kanamori}.
The Hamiltonian matrix is given in Eq.~\eqref{eq:HMF}, with eigenvalues and eigenvectors defined in the main text.
In its turn, the mean-field free energy reads as Eq.~\eqref{eq:FMF_Kanamori}.
To derive the jacobian, necessary to efficiently minimize the grand potential, we compute derivatives of $F$ with respect to $M_{\mu}$ and $\phi_{\mu}$.
We use the identity
\begin{align}
\frac{\partial}{\partial \lambda}
\left(
-\frac{1}{\beta}
\sum_{\boldsymbol{k}\nu}
\ln(1+e^{-\beta \epsilon_{\boldsymbol{k}\nu}})
\right)
=
\sum_{\boldsymbol{k}\nu}
f(\epsilon_{\boldsymbol{k}\nu})
\frac{\partial \epsilon_{\boldsymbol{k}\nu}}{\partial \lambda},
\label{eq:free_energy_derivative_identity}
\end{align}
where $f(E)=1/(e^{\beta E}+1)$ is the Fermi function.
Using the Hellmann-Feynman theorem,
\begin{align}
\frac{\partial \epsilon_{\boldsymbol{k}\nu}}{\partial \lambda}
=
\left\langle \boldsymbol{k} \nu \sigma
\left|
\frac{\partial \hat{h}^{(\sigma)}(\boldsymbol{k})}{\partial \lambda}
\right|
\boldsymbol{k} \nu \sigma
\right\rangle.
\label{eq:HF_theorem}
\end{align}

We compute
\begin{align}
\frac{\partial F}{\partial M_{\mu}}.
\end{align}
The bosonic contribution gives
\begin{align}
\frac{\partial}{\partial M_{\mu}}
\left(
\frac{1}{2}
\sum_{\eta\alpha}
M_{\eta}
(U_s^{-1})^{\eta\eta}_{\alpha\alpha}
M_{\alpha}
\right)
=
\sum_{\eta}
(U_s^{-1})^{\mu\mu}_{\eta\eta}
M_{\eta}.
\label{eq:bosonic_derivative_M}
\end{align}
The fermionic contribution yields
\begin{align}
\sum_{\boldsymbol{k} \nu \sigma}
f(\epsilon_{\boldsymbol{k}\nu}-\mu)
\left\langle \boldsymbol{k} \nu \sigma
\left|
\frac{\partial \hat{h}^{(\sigma)}(\boldsymbol{k})}{\partial M_{\mu}}
\right|
\boldsymbol{k} \nu \sigma
\right\rangle.
\end{align}
We define
\begin{align}
\hat{\Delta}^{\mu}
=
-\sigma
\frac{\partial \hat{h}^{(\sigma)}(\boldsymbol{k})}{\partial M_{\mu}}=
\begin{pmatrix}
0 & \hat{1}_{\mu} \\
\hat 1_{\mu} & 0
\end{pmatrix},\label{eq:delta}
\end{align}
where $\hat 1_{\mu}$ is 1 in the diagonal orbital matrix element $\mu$ and zero in all the other matrix elements.
The full derivative reads
\begin{equation}
\begin{split}
\frac{\partial F}{\partial M_{\mu}}
&=
\sum_{\eta}
(U_s^{-1})^{\mu\mu}_{\eta\eta}
M_{\eta}\\
&-
\frac{1}{N}
\sum_{\boldsymbol{k}\in RBZ}
\sum_{\nu\sigma}
\sigma
\langle \boldsymbol{k} \nu \sigma |
\hat{\Delta}^{\mu}
| \boldsymbol{k} \nu \sigma \rangle
f(\epsilon_{\boldsymbol{k}\nu}-\mu).
\end{split}
\label{eq:gap_mag}
\end{equation}

Similarly, we compute
\begin{align}
\frac{\partial F}{\partial \phi_{\mu}}.
\end{align}
The bosonic part gives
\begin{align}
\frac{\partial}{\partial \phi_{\mu}}
\left(
\frac{1}{2}
\sum_{\eta\alpha}
\phi_{\eta}
(U_c^{-1})^{\eta\eta}_{\alpha\alpha}
\phi_{\alpha}
\right)
=
\sum_{\eta}
(U_c^{-1})^{\mu\mu}_{\eta\eta}
\phi_{\eta}.
\end{align}
The fermionic part gives
\begin{align}
\sum_{\boldsymbol{k} \nu \sigma}
f(\epsilon_{\boldsymbol{k}\nu}-\mu)
\left\langle \boldsymbol{k} \nu \sigma
\left|
\frac{\partial \hat{h}^{(\sigma)}(\boldsymbol{k})}{\partial \phi_{\mu}}
\right|
\boldsymbol{k} \nu \sigma
\right\rangle.
\end{align}
We define
\begin{align}
\hat{N}^{\mu}
=
\frac{\partial \hat{h}^{(\sigma)}(\boldsymbol{k})}{\partial \phi_{\mu}}=
\begin{pmatrix}
\hat{1}_{\mu} & 0 \\
0 & \hat 1_{\mu}
\end{pmatrix}.
\end{align}
Hence,
\begin{equation}
\begin{split}
\frac{\partial F}{\partial \phi_{\mu}}
&=
\sum_{\eta}
(U_c^{-1})^{\mu\mu}_{\eta\eta}
\phi_{\eta}\\
&+
\frac{1}{N}
\sum_{\boldsymbol{k}\in RBZ}
\sum_{\nu\sigma}
\langle \boldsymbol{k} \nu \sigma |
\hat{N}^{\mu}
| \boldsymbol{k} \nu \sigma \rangle
f(\epsilon_{\boldsymbol{k}\nu}-\mu).
\end{split}
\label{eq:gap_charge}
\end{equation}

The chemical potential $\mu$ must be determined self-consistently in order to fix the total electron density. 
It enters the mean-field Hamiltonian through
\begin{align}
\xi_{\boldsymbol{k}}^{\mu\eta}
=
\epsilon_{\boldsymbol{k}}^{\mu\eta}
-
\mu\delta_{\mu\eta}.
\end{align}
Since the grand potential $\Omega=F+\mu n$, minimization on $\mu$ implies
\begin{align}
n
=
-\,\frac{\partial F}{\partial \mu}.
\end{align}
Using
\begin{align}
\frac{\partial \epsilon_{\boldsymbol{k}\nu}}{\partial \mu}
=
\left\langle \boldsymbol{k} \nu \sigma
\left|
\frac{\partial \hat{h}^{(\sigma)}(\boldsymbol{k})}{\partial \mu}
\right|
\boldsymbol{k} \nu \sigma
\right\rangle=
-{1},
\end{align}
we obtain
\begin{align}
-\frac{\partial F}{\partial \mu}
=
\frac{1}{N}
\sum_{\boldsymbol{k}\in RBZ}
\sum_{\nu\sigma}
f(\epsilon_{\boldsymbol{k}\nu}).
\end{align}
Therefore, the chemical potential equation fixing the filling $n$ is
\begin{align}
n
=
\frac{1}{N}
\sum_{\boldsymbol{k}\in RBZ}
\sum_{\nu\sigma}
f(\epsilon_{\boldsymbol{k}\nu}).\label{eq:mu_kanamori}
\end{align}
Eqs.~\eqref{eq:gap_mag} and \eqref{eq:gap_charge} are provided to the minimization algorithm and \eqref{eq:mu_kanamori} is solved implicitly at every iteration of the minimization procedure, self-consistently for all $M_{\mu}$, $\phi_{\mu}$, and $\mu$.

\begin{figure}[t]
\begin{center}
\includegraphics[width=1.0\columnwidth]{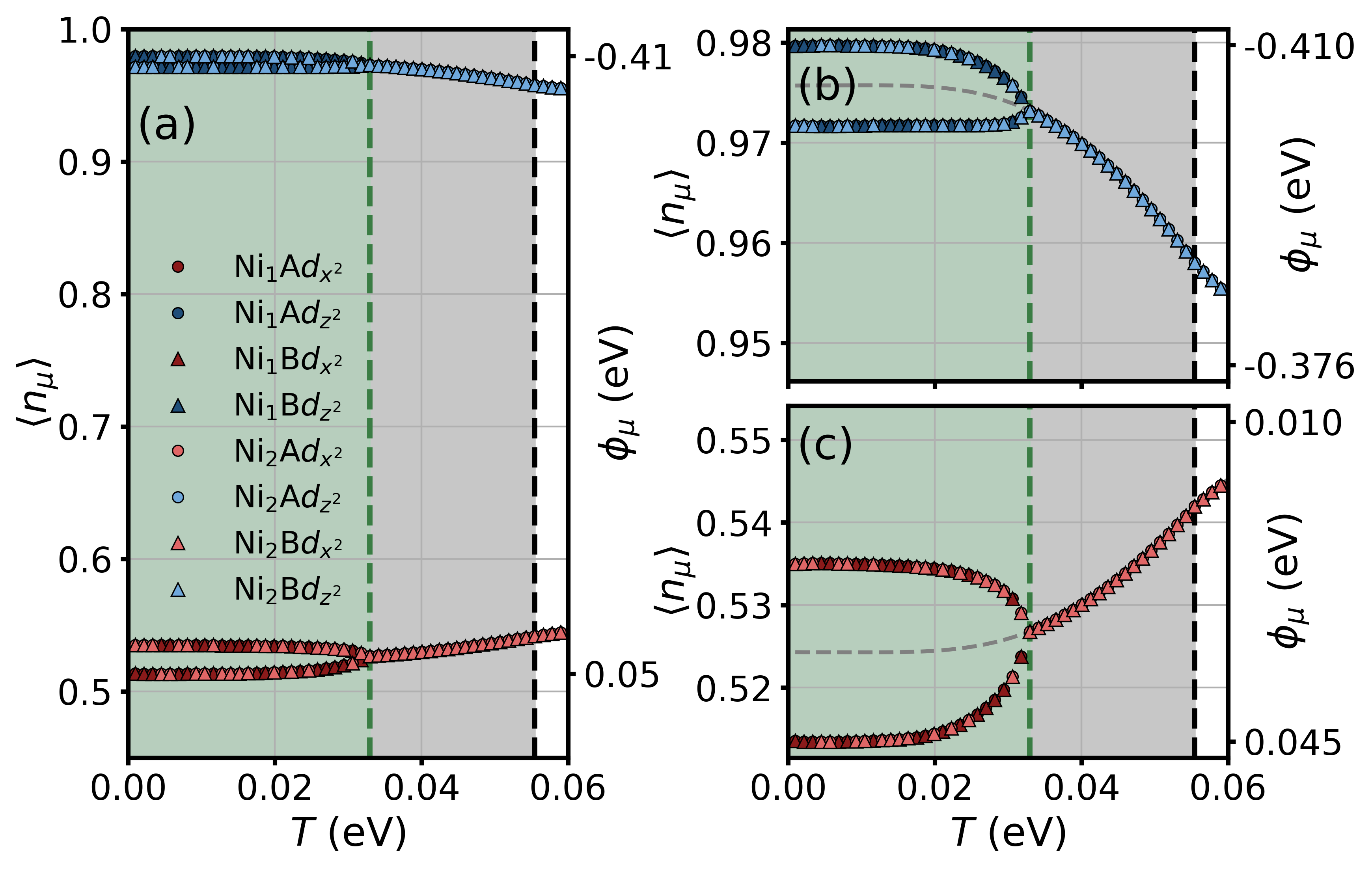}
\caption{ Charge densities as a function of temperature for the data shown in the main text Fig.~\ref{fig:order_parameters}.
Panel (a) show all the eight orbitals and zooms are shown for the $d_{3z^2-r^2}$ (b) and $d_{x^2-y^2}$ (c) orbital-resolved densities.
Gray lines in panels (b) and (c) show averages over all four sites for each orbital.
Noticeably, $T_{\text{SDW}}$ does not affect the densities, whereas $T_{\text{DW}}$ induces the splittings shown in the main text.
Circles lie on top of triangle symbols of the same color, implying layer equivalence.
}
\label{fig:charge}
\end{center}
\end{figure}

\section{Order parameters of low-pressure La$_3$Ni$_2$O$_7$}
\label{appendix:details}
In this appendix, we give more details on the order parameters as a function of temperature and interaction strengths $U$ and $J_H$.
The analysis of the order parameters allows an experiment-motivated choice of $U$ and $J_H$.

Fig. \ref{fig:charge} shows more details on the charge order presented in the main text.
At $T<T_{\text{DW}}$, spins in the diagonal sites in e.g. layer A of Ni$_1$ and layer B of Ni$_2$ become nonequivalent, as shown in Fig.~\ref{fig:order_parameters}(a,b).
This second symmetry-breaking is accompanied by a charge disproportionation at the same two sites, and is shown in better details for all orbital components in Fig.~\ref{fig:charge}(a) and zoomed in panels (b) and (c).

\begin{figure}[t]
\begin{center}
\includegraphics[width=1.0\columnwidth]{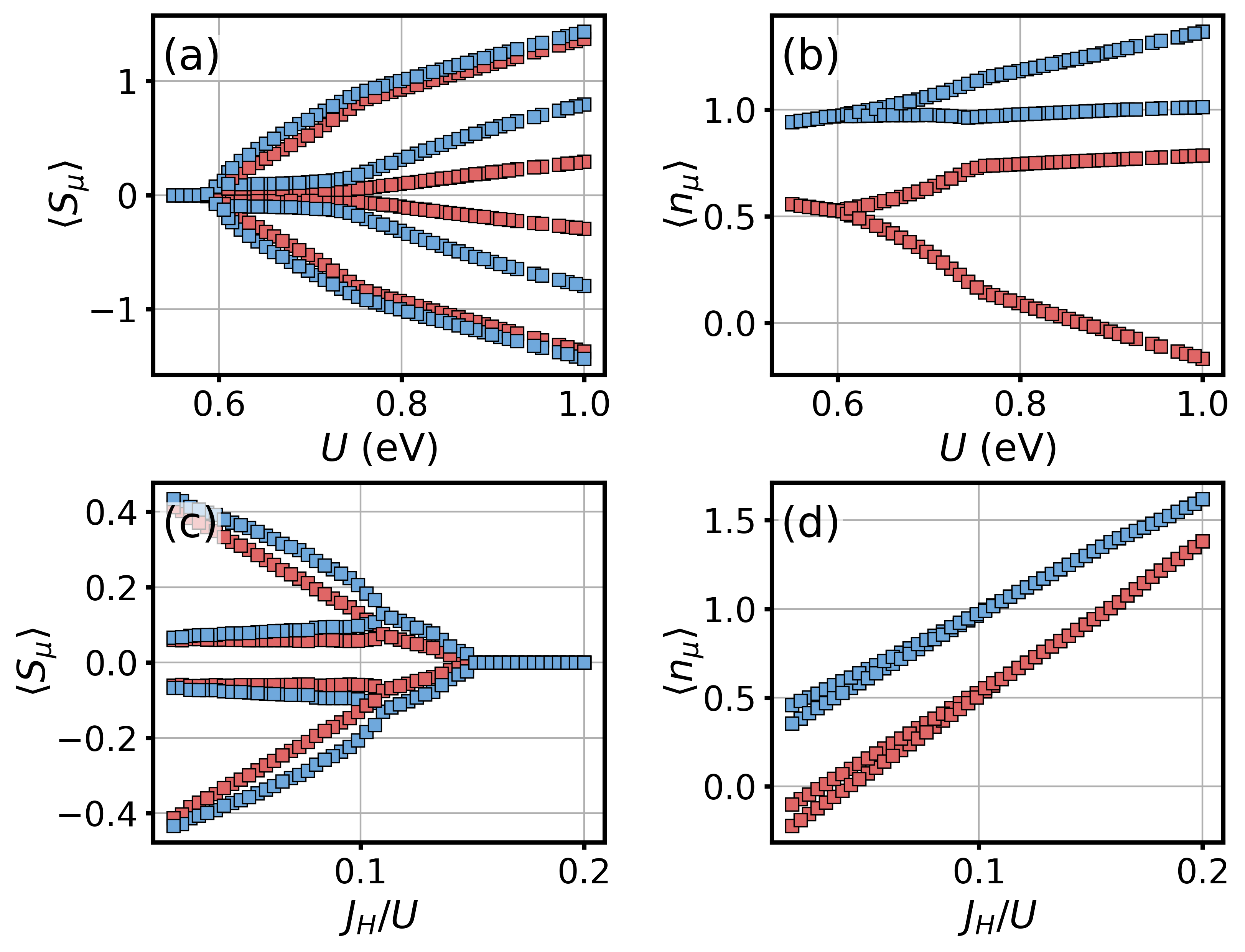}
\caption{ Onsite Hubbard $U$ [(a) and (b)] and Hund's $J_H$ interaction dependence of the spin $\langle S_{\mu}\rangle$ and charge $\langle n_\mu\rangle$ densities.
Blue and red colors indicate $d_{3z^2-r^2}$ and $d_{x^2-y^2}$ orbitals, respectively.
Panels (c) and (d) use constant $U=0.61$ eV.
In these simulations, $T=0.001$ eV.
}
\label{fig:interactions}
\end{center}
\end{figure}

\begin{figure}[t]
\begin{center}
\includegraphics[width=1.0\columnwidth]{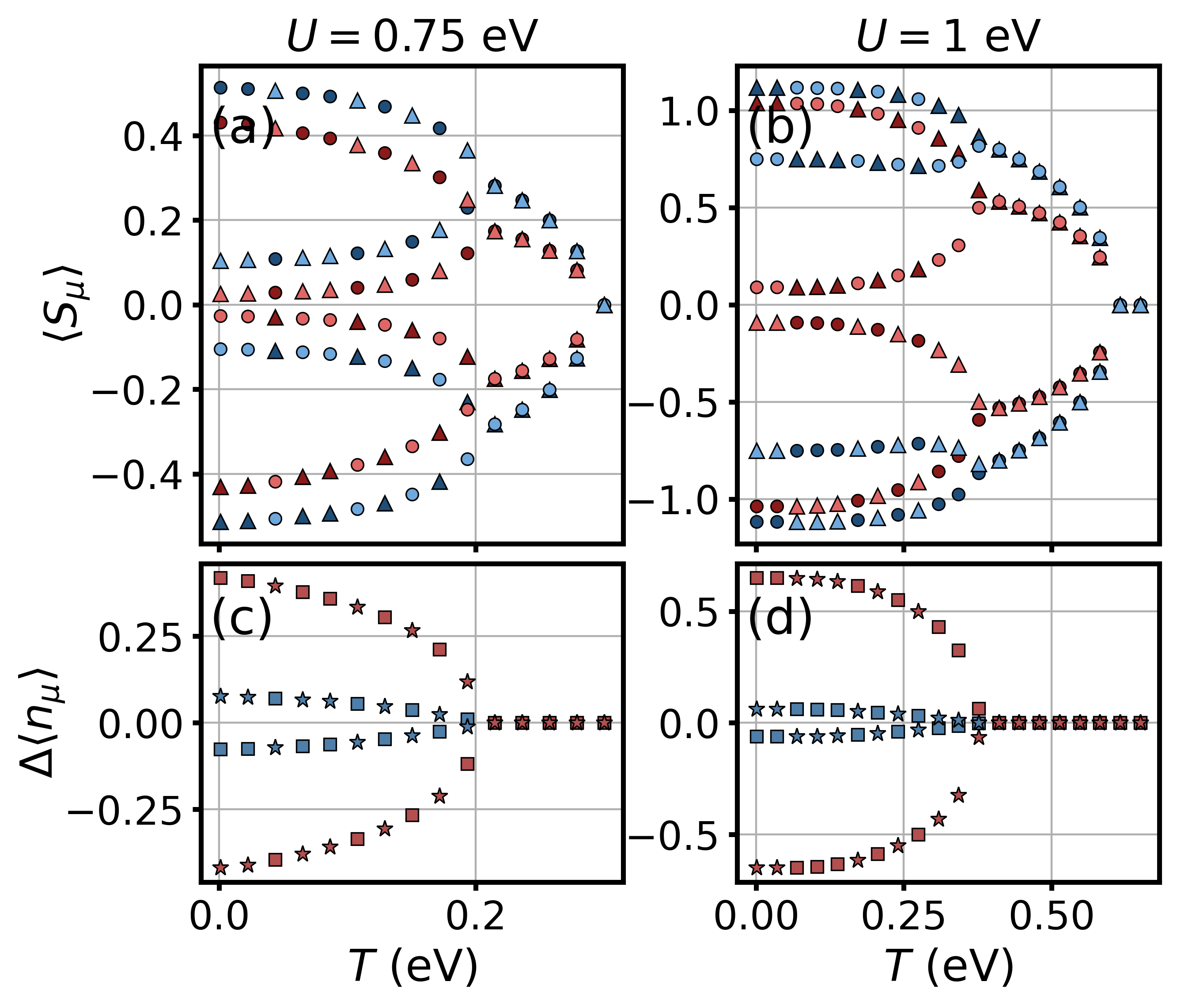}
\caption{ {Robustness of the $T_{\text{SDW}}>T_{\text{DW}}$ hierarchy. 
Same as Fig.~\ref{fig:order_parameters} for $U=0.75$ eV (left column) and $U=1$ eV (right column) with $J=U/10$.
Symbols adopt the same convention used in Fig.~\ref{fig:order_parameters}.}
}
\label{fig:interactions_T}
\end{center}
\end{figure}

We show the dependence of our results on $U$ and $J_H$ in Fig.~\ref{fig:interactions}.
Increasing $U$ [panels (a) and (b)] makes the effects probed here more pronounced.
We remark the larger scale of $\langle S_\mu\rangle$ as compared to the main text Fig.~\ref{fig:order_parameters}(a).
$U>1$ eV fully gaps the Fermi level, which is not supported by any of the available experimental probes for the present system.
As shown in panel (c), lowering $J_H$ from $J_H=U/10$ increases the magnetic order parameter.
Additionally, lower Hund's coupling $J_H$ makes the splitting between orbital charge densities more pronounced [panel (d)]. In contrast, increasing $J_H$ first suppresses the additional splitting and finally also the double stripe phase completely. The scale of the order parameters are used to constraint the $U$ and $J_H$ values to $\max M_\mu\sim100$ meV, which is accomplished for $U=0.61$ eV and $J_H=U/10$.
{Finally, Fig.~\ref{fig:interactions_T} shows the temperature evolution of the spin and charge expectation values for $U=0.75$ eV (left column) and $U=1$ eV (right column) keeping $J=U/10$, where in the latter case the SDW gap is large enough to gap the Fermi surface out at lower temperatures.
The results shown in Fig.~\ref{fig:interactions_T} corroborate for the robustness of the $T_{\text{SDW}}>T_{\text{DW}}$ hierarchy in our model across a range of Hubbard $U$ values.}

\bibliography{references}

\end{document}